\documentclass[structabstract]{aa}  
\usepackage{graphicx}
\usepackage{txfonts}
\def\Ks{\hbox{$K$s}}

\def\Js{\hbox{$J$s}}
\def\J{\hbox{$J$}}
\def\H{\hbox{$H$}}
\newcommand{\Htwo}{{\hbox {\ensuremath{\mathrm{H_2}}}}}

\newcommand{\twco}{{\hbox {\ensuremath{\mathrm{^{12}CO}} }}}
\newcommand{\twcop}{{\hbox {\ensuremath{\mathrm{^{12}CO}}}}}

\newcommand{\thco}{{\hbox {\ensuremath{\mathrm{^{13}CO}}}}}
\newcommand{\thcop}{{\hbox {\ensuremath{\mathrm{^{13}CO}}}}}
\newcommand{\kmps}{\ensuremath{\mathrm{km\,s^{-1}}}}

\begin{document}

\title{Mass and motion of globulettes in the Rosette Nebula   \thanks{Based on observations collected at Onsala Space Observatory, Sweden, European Southern Observatory, Chile (084.C-0299(A) and 088.C-0630(A)) and Nordic Optical Telescope, La Palma, Spain, and with the Atacama Pathfinder Experiment (APEX), Llano Chajnantor, Chile (O-088.F-9318A)}}


  \author{ G. F. Gahm\inst{1}
          \and C. M. Persson\inst{2}
	        \and M. M. M\"akel\"a\inst{3}
	        \and L. K. Haikala\inst{3,4}}

   \offprints{G. F. Gahm}

   \institute{Stockholm Observatory, AlbaNova University Centre, Stockholm University,
	      SE-106 91 Stockholm, Sweden\\
              email: \mbox{gahm@astro.su.se}
              \and Chalmers University of Technology, Department of Earth and Space Sciences, Onsala Space Observatory, SE-439 92 Onsala, Sweden 
              \and Department of Physics, PO Box 64, FI-00014 University of Helsinki, Finland    
              \and Finnish Centre for  Astronomy with ESO (FINCA), University of Turku, V\"ais\"al\"antie 20, FI-21500  Piikki\"o, Finland}

   \date{}

 \abstract
 {Tiny molecular clumps are abundant in many \ion{H}{ii} regions surrounding newborn stellar clusters. In optical images these so-called globulettes appear as dark patches against the background of bright nebulosity.}
{ We aim to clarify the physical nature of globulettes in the Rosette Nebula by deriving densities and masses, and to determine their velocities as a function of position over the nebula.}
 { Radio observations were made of molecular line emission from 16 globulettes identified in a previous optical survey.  In addtion, we collected images in the NIR broad-band JHKs and narrow-band Paschen $\beta$ and H2. Ten objects, for which we collected information from several transitions in 12CO and 13CO were modelled using a spherically symmetric model.}
 { Practically all globulettes were detected in our CO survey. The best fit to observed line ratios and intensities was obtained by assuming a model composed of a cool and dense centre and warm and dense surface layer. The average masses derived range from about 50 to 500 Jupiter masses, which is similar to earlier estimates based on extinction measures. The globulettes selected are dense, with very thin layers of fluorescent H2 emission. The NIR data shows that several globulettes are very opaque and contain dense cores. }
 { We conclude that the entire complex of shells, elephant trunks, and globulettes in the northern part of the nebula is expanding with nearly the same velocity of ~22 km/s, and with a very small spread in velocity among the globulettes. Some globulettes are in the process of detaching from elephant trunks and shells, while other more isolated objects must have detached long ago and are lagging behind in the general expansion of the molecular shell. The suggestion that some globulettes might collapse to form planetary-mass objects or brown dwarfs is strengthened by our finding of dense cores in several objects. }


\maketitle
%

\section{Introduction}
\label{sec:intro}

Young stellar clusters are surrounded by expanding \ion{H}{ii} regions as a  
result of light and winds from central O and B stars that act on the ambient molecular  
cloud. Optical images of such magnificent nebulae show a mix of bright and dark  
nebulosity. The dark features mark remnants of the cold molecular cloud that have 
been compressed into shells and accelerated away from the central cluster. They 
contain dust, and are seen in silhouette against the bright background coming 
mainly from emission lines formed in the ionized bubble. Remarkable formations 
appear in these shells, such as elongated dusty pillars, the so-called elephant trunks,  
that point like fingers to the central cluster.

A number of such \ion{H}{ii} regions contain distinct, but very small clumps, as  
recognized early by Minkowski (\cite{minkowski49}) and Thakeray (\cite{thakeray50}).  
Herbig (\cite{herbig74}) noted that many such cloudlets found in the Rosette Nebula have 
teardrop forms with bright rims facing the central cluster. More detailed studies followed, and several of these  
focused on the so-called proplyds, which are photoevaporating disks surrounding very  
young stars (Bally et~al. \cite{bally00};   Smith et~al. \cite{smith03}, and 
references  therein). In these studies small cloudlets without any obvious central  
stellar objects were recognized as well.

Hester et~al. (\cite{hester96}) drew the attention to some very tiny (photo-)
evaporating gas globules   (EGGs) in the Eagle Nebula, without any sign of embedded stars. 
Since then, such starless cloudlets have been searched for in more than 20 \ion{H}{ii} regions  in  
systematic studies in Reipurth et~al. (\cite{reipurth97, reipurth03}), De Marco et~al. 
(\cite{marco06}), Gahm et~al. (\cite{gahm07}, hereafter called Paper~I), and Ito et~al. 
(\cite{ito08}). These studies show that the majority of these objects have radii 
$<$10~kAU with size distributions that typically peak at $\sim$2.5~kAU. In Paper~I,  
masses were derived from extinction measures, indicating that most objects have masses  
$<$13~M$_{J}$ (Jupiter masses), which currently is taken to be the domain of 
planetary-mass objects. Hence, these tiny clouds form a distinct class of objects and  
are unrelated to the much larger globules spread in interstellar space. The objects   
have been given many names, but here we stick to the designation proposed in Paper~I,  
namely {\it globulettes}, to distinguish them from globules  and proplyds. 

Most globulettes appear as dark, roundish patches in H$\alpha$ images and show no traces  
of bright rims. Some objects are more elongated and even teardrop shaped. Only a small   
fraction of the globulettes have distinct bright rims or halos. Many objects are quite  
isolated and located far from molecular shells and elephant trunks in the regions.  
Others are seen clustering close to trunks or shells, and there are objects connected 
by thin filaments to the larger blocks, suggesting that globulettes may form as a  
consequence of erosion of these larger structures.  

The fact that many globulettes are quite isolated from larger molecular blocks 
indicates that they can survive for quite some time in harsh environments. This is  
in accordance with the long survival times derived in Paper~I, based on the analytic  
treatment of photoevaporating clumps by Mellema et~al. (\cite{mellema98}), and from 
more detailed 3D numerical simulations by Kuutmann (\cite{kuutmann07}). In the latter  
study lifetimes of $\sim$5$\times$10$^4$ years were obtained for $\sim$30~$M_{J}$ 
objects, and the expected lifetime increases with mass. In these models, the globulettes are protected  
against immediate photodisintegration by an ionized bubble that expands like an 
umbrella in the direction of the cluster, and interacts with and removes impinging UV photons.

In the numerical simulations the objects develop bright rims from the ionized gas on 
the side facing the clusters. Moreover, the objects take teardrop forms, and eventually  
very long, dusty tails extend from the backside, in the opposite direction from the 
cluster. The cores are also accelerated through the interaction with light from 
central stars, known as the "rocket effect". In the calculations by 
Kuutmann  (\cite{kuutmann07}) a globulette  can reach terminal velocities 
of  5 to 10~km~s$^{-1}$ with respect to the initial rest frame. However, one expects 
that the globulettes move outwards from the start with the same velocity as 
the system of trunks and shells. 

The outer pressure from the surrounding warm plasma and the radiation pressure 
contribute in confining the objects, and one expects that the initial penetrating 
shock, generated by photoionization, leads to compression of the interior.  
As discussed in Paper~I, some of the larger globulettes 
may already have, or can develop, denser cores that collapse to form brown dwarfs, 
or planetary-mass objects, before evaporation has proceeded very far. Such objects 
initially have high velocities in the direction away from the cluster and will therefore eventually 
shoot out to the galactic environment like interstellar bullets. 

Many low mass objects in interstellar space have been confirmed to be free-floating 
planetary-mass objects (Quanz et~al. \cite{quanz10}, and references therein). 
From  a more recent survey of sources of gravitational microlensing, Sumi et~al.  
(\cite{sumi11}) concluded that the total number of unbound planets in the Milky Way  
could be very large indeed, several hundred billions, or almost twice the number 
of main-sequence stars. It is usually assumed that such planets once formed in  
circumstellar protoplanetary disks, where perturbations led to ejection 
(e.g. Veras et~al. \cite{veras09}). Since there are binaries composed of very 
low-mass objects  (e.g. Jayawardhana et~al. \cite{jaya06}; Gelino et al. \cite{gelino11}; 
Burgasser et al. \cite{burg12}), it is clear that  isolated low-mass objects also form $\it {in~situ}$ 
from isolated small cloudlets. In Paper~I it was suggested that globulettes in \ion{H}{ii} regions could be an  
alternative source of feeding the galaxy with free-floating planetary-mass objects.

There are many questions concerning the nature and fate of globulettes, and it 
is desirable to collect more information on their physical state. A great puzzle 
is that the majority show no signs of bright rims, teardrop forms, or extended dusty 
tails, as predicted by the models. Inclination is obviously of importance, since  
elongated objects may appear round if they are oriented along the line-of-sight. 
As a consequence, bright rims on the remote side could escape detection since they  
suffer from extinction in the optical spectral region. Another question is whether 
some globulettes have denser cores or even host stars or planets. 
The masses derived in Paper~I were based on column densities of dust and an 
assumed gas-to-dust ratio of 100. Complementary molecular line observations 
would provide  independent mass estimates based on the gas content. 
Finally, it is of interest to find out how the globulettes move relative to 
adjacent trunks and shells, and if there is evidence that they have been 
accelerated due to the "rocket effect".

We have conducted radio observations of molecular line emission and near-infrared
(NIR) broad and narrow-band imaging to shed more light on the properties of globulettes. 
We have selected objects in the \object{Rosette Nebula} surrounding the cluster \object{NGC 2244} from the sample in Paper~I, and added some cloudlets of special interest in the same area. 

For information on previous studies of the Rosette Nebula, including NIR surveys, see the review by R\'oman-Z\'u\~{n}iga (\cite{roman08}). Of particular relevance to our molecular line observations is the large-scale CO survey by Dent et~al. (\cite{dent09}) and the maps in CO and CS of cloudlets by Gonz\'alez-Alfonso \& Cernicharo (\cite{gonzalez94}). Elephant trunks in the area have been mapped by Schneps et~al. (\cite{schneps80}) and Gahm et~al. (\cite{gahm06}).

\section{Observations, reductions, and objects}
\label{sec:obs}

\subsection{Observations at Onsala space observatory}
\label{sec:oso}

The  observations were carried out in April to June 2010, February 
 to March 2011, and February 2012 using a 3-mm   SIS receiver at the 20-m telescope at  Onsala Space Observatory (OSO).  We observed $^{12}$CO  \mbox{$J$\,=\,1\,--\,0}  at 115.271~GHz, and/or $^{13}$CO  \mbox{$J$\,=\,1\,--\,0}  at 110.201~GHz. 
A few objects were also observed in the HCO$^{+}$ $J$\,=\,1\,--\,0 transition at 89.189~GHz. 
All globulettes and observed transitions in the respective object are  listed in Table~\ref{table:objects}.

The observations were performed in the frequency-switching mode, using a 1600-channel hybrid
digital autocorrelation spectrometer  at 25 kHz channel spacing ($\Delta v\!\sim\! 0.07$~km~s$^{-1}$)
and a bandwidth of 40~MHz. The single-sideband system temperature was typically around \mbox{1\,000\,--\,1\,400~K} for $^{12}$CO, and \mbox{400\,--\,600~K} for $^{13}$CO. The pointing was checked regularly, and we estimate the pointing error to be less than a few arcseconds. At 115~GHz, the full width at half maximum (FWHM)  beam size of the 20 m-antenna is 33$\arcsec$, and the main beam efficiency $\sim$ 0.30 (for an average elevation of approximately 30$^\circ$). The chopper-wheel method was used for the intensity calibration. The data reduction was performed with the spectral line software package  {\tt xs}\footnote{{\tt http://www.chalmers.se/rss/oso-en/observations/data-\\reduction-software}}.

To ensure that the observed signals originate in the selected globulette off-set positions, normally at 30$\arcsec$ or 45$\arcsec$ from the globulette, were also observed. Some globulettes are projected on a background of more extended molecular emission, and for these, and also for a few objects with complex line profiles, additional off-set  positions were included. 

\subsection{APEX observations} \label{subsection: Apex observations}
\label{sec:apex}

Observations of two higher $^{12}$CO and  $^{13}$CO transitions were carried out 
in September and November 2011 with the 12~m APEX{\footnote {This publication is partly based on data acquired with the Atacama Pathfinder Experiment (APEX). APEX is a collaboration between the Max-Planck-Institut f\"ur Radioastronomie, the European Southern Observatory, and the Onsala Space Observatory.}} telescope at Llano Chajnantor, Chile: 
$^{12}$CO \mbox{$J$\,=\,(2\,--\,1)} at 230.538~GHz, and \mbox{$J$\,=\,(3\,--\,2)} at 345.796~GHz, and
$^{13}$CO \mbox{$J$\,=\,(2\,--\,1)} at 220.399~GHz, and 
\mbox{$J$\,=\,(3\,--\,2)} at 330.588~GHz.  We used two single sideband heterodyne SIS-receivers mounted on 
the Nasmyth-A focus: APEX-1 and APEX-2. All observations were performed in 
position-switching mode. The FWHM is 27$\arcsec$ at 230 GHz, and 18$\arcsec$ at 345 GHz, 
and the corresponding main beam efficiencies are 0.75 and 0.73.
The RPG eXtended bandwidth Fast Fourier Transform Spectrometer (XFFTS)
has 32\,768 channels and consists of two units with  a fixed overlap 
region of 1.0~GHz and an instantaneous bandwidth of 2.5 GHz. 
The channel spacing is thus 76.3~kHz, and the velocity spacing
$\Delta v$ is 0.1 and 0.07~km\,s$^{-1}$ at 230 and 345~GHz, respectively. 
The  pointing was checked regularly, and the error is estimated
to be within 2$\arcsec$. The data reduction of the APEX data was performed with CLASS, part of the 
GILDAS-package\footnote{http://iram.fr/IRAMFR/GILDAS/}, and also checked with {\tt xs}.

Globulettes of different sizes were selected from the sample observed at 
OSO (see Table~\ref{table:objects}), and a few off-set positions were included in the APEX observations as well to confirm the OSO results.

\subsection{NIR observations}
 \label{subsection:NTT}

\begin{table*}[!ht] 
\centering 
\caption[]{Globulettes and shells observed with the Onsala 20-m antenna, APEX, and NTT.} 
\begin{tabular}{lccccl | lccccl} 

\hline\hline
     \noalign{\smallskip}
RN\tablefootmark{a}& \multicolumn{2}{c}{OSO\tablefootmark{b}}  & APEX\tablefootmark{c} & NTT & Notes  &
RN\tablefootmark{a}& \multicolumn{2}{c}{OSO\tablefootmark{b}}& APEX\tablefootmark{c} & NTT & Notes  \\ 
&  $^{12}$CO  & $^{13}$CO  &&&& & $^{12}$CO  & $^{13}$CO \\
     \noalign{\smallskip}
     \hline
       
5 & Map & x & x  &  & HCO$^{+}$ , D 10 & A&  & Map & Map &   &ACIS \#86, D 8 \\
9& x & x & x &  &  & B&  & x &  &  &\\
12&  & x & &  &  & C& x & x &  &  & HCO$^{+}$ \\
31& x & x &  & x &   & D&  & x &  &  & D15\\
35&  & x &  & x &  & Shell A& x &  &  &  &\\
38& x & x & x & x & D 6   & Shell B& x &  & & x & \\
40& x  & x & x  & x & HCO$^{+}$, TDR 10, D 6  & Shell C&  & x &  & x &\\
63& x  & x &  &  &  & Claw E& x &   &  & x &  \\
88&  & x & x & x & & Shell D& x &  &  &  & \\
93(+91)&  & x & x & x &&&\\
95(+94)& x & x & x & x  &&&\\
101&  & x &  &  &&&&\\
110& x & x & &  &&&&\\
114&  & x & x & x &&&&\\
122& x & x & x & x &&&&\\
129& x & x & x & x & HCO$^{+}$&&& \\
\hline
    \noalign{\smallskip}

\label{Table: transitions}
\end{tabular}
\tablefoot{
\tablefoottext{a}{Central positions, RA and Dec. ($J$2000.0), for objects not listed in Paper~I. 
A: 06:30:50.3 +05:00:28.7; B: 06:31:29 +05:16:16;  
C: 06:31:42.4 +05:22:21; D: 06:31:44 +05:03:12;  
Shell A: 06:30:49.5 +05:00:45; Shell B: 06:30:52.3 +05:06:26.4; 
Shell C: 06:31:07.5 +05:07:35; Claw E: 06:31:39.5 +05:11:36; 
Shell D: 06:32:09 +05:16:54}
\tablefoottext{b}{$J$\,=\,1\,--\,0.}
\tablefoottext{c}{$^{12}$CO and $^{13}$CO, $J$\,=\,2\,--\,1 and 3\,--\,2.}
}
\label{table:objects} 
\end{table*} 

Four fields containing radio-detected globulettes in the Rosette Nebula were imaged with the Son of Isaac
(SOFI) infrared spectrograph and imaging camera on the New Technology Telescope (NTT) at the European Southern Observatory in Chile using the  \J, \H, and \Ks\ broad-band filters, and the narrow filters NB H2 S1, which covers the \Htwo\ 1-0 S(1) 2.12 $\mu$m spectral line, and NB 2.090, which covers the adjacent continuum. The observations were conducted during four nights in December 2009 though January 2010.  The SOFI field of view is 4\farcm9, and its pixel size 0\farcs288. The $\J\H\Ks$ observations were carried out in the on-off mode instead of in the standard jitter mode to retain the possible extended surface brightness features. After each on-integration an off-position outside the Rosette Nebula was observed, and jittering was performed after two on-off pairs. One-minute integration time, which consisted of six individual 10-second integrations, was used in \J\H\Ks. The observed on-source times per  filter were 26 min, except for one field (Field 19), which was observed for 13 minutes.
 
The \Htwo\ observations were obtained in the jitter mode, where the surface brightness with a scale larger than the jitter box (30\arcsec) is smeared and/or cancelled in the data reduction. Only small-size features and gradients in the original surface brightness structure were retained. Point-like sources and galaxies were unaffected by the jittering. The fields have 40 frames of one-minute integrations obtained using three 20- or two  30-second integrations. The average seeing during the observations was $\sim$ 0\farcs8. Because the NB 2.090 is narrower than the NB H2 S1 filter, a 20\%\ longer total integration was required in this filter. In addition to the broad-band on-off observations, one SOFI field (Field 19) was observed in the jitter mode in \Js\H\Ks. Complementary SOFI imaging of several fields was performed in January 2012 through a narrow-band Paschen $\beta$ filter (at 1.28 $\mu$m) combined with a narrow-band continuum filter.  

The IRAF{\footnote {IRAF is distributed by the National Optical Astronomy Observatories, which are operated by the Association of Universities for Research in Astronomy, Inc., under cooperative agreement with the National Science Foundations}} external XDIMSUM package was used in the data reduction. The images were searched for cosmic rays, were sky-subtracted, flat-fielded, illumination-corrected, registered and averaged. For the $\J\H\Ks$ observations the four off-position images nearest in time to each on-position image were used in the sky-subtraction. For the jittered images the two neighbouring images were used. An object mask was constructed for each off image. Applying these masks in the sky-subtraction produced hole-masks for each sky-subtracted image.  Special dome flats and illumination-correction frames provided by the NTT team were used to flat-field and to illumination-correct the sky-subtracted images. Rejection masks combined from a bad-pixel mask and individual cosmic-ray and hole masks were used when averaging the registered images.

The Source Extractor software v.2.5.0 (Bertin \& Arnouts \cite{bertin96}) was used to extract the photometry for each reduced SOFI field. The steps to build the stellar photometry catalogues out of the SOFI magnitudes are the same as in M{\"a}kel{\"a} \& Haikala (\cite{makela13}). Objects within 30 pixels from the frame edges and objects with a Source Extractor star index less than 0.9 in more than one band were discarded. After combining the catalogues for each field, 1375 objects remain. The limiting magnitudes for a formal error of 0\fm15 are \J\ $\approx$ 21, \H\ $\approx$ 20, \Ks\ $\approx$ 19.5 except for Field 19, where they are 0\fm5 brighter.

\subsection{Selected objects}
\label{sec:objects}

The globulettes investigated in Paper~I were identified from deep narrow-band H$\alpha$ 
images collected with the $2.6$ m Nordic Optical Telescope (NOT) on La Palma, Canary 
Islands, Spain. In the Rosette Nebula a total of 145 globulettes were listed. We have 
observed 16 of these at OSO and 10 with APEX. The objects were selected as to cover such broad a range in mass as possible, but objects with masses in the planetary domain could not be reached. Moreover, the selected objects are relatively dense in comparison with some more diffuse globulettes listed in Paper~I.

\begin{figure*}[t]
\centering
\includegraphics[angle=00, width=18.5cm]{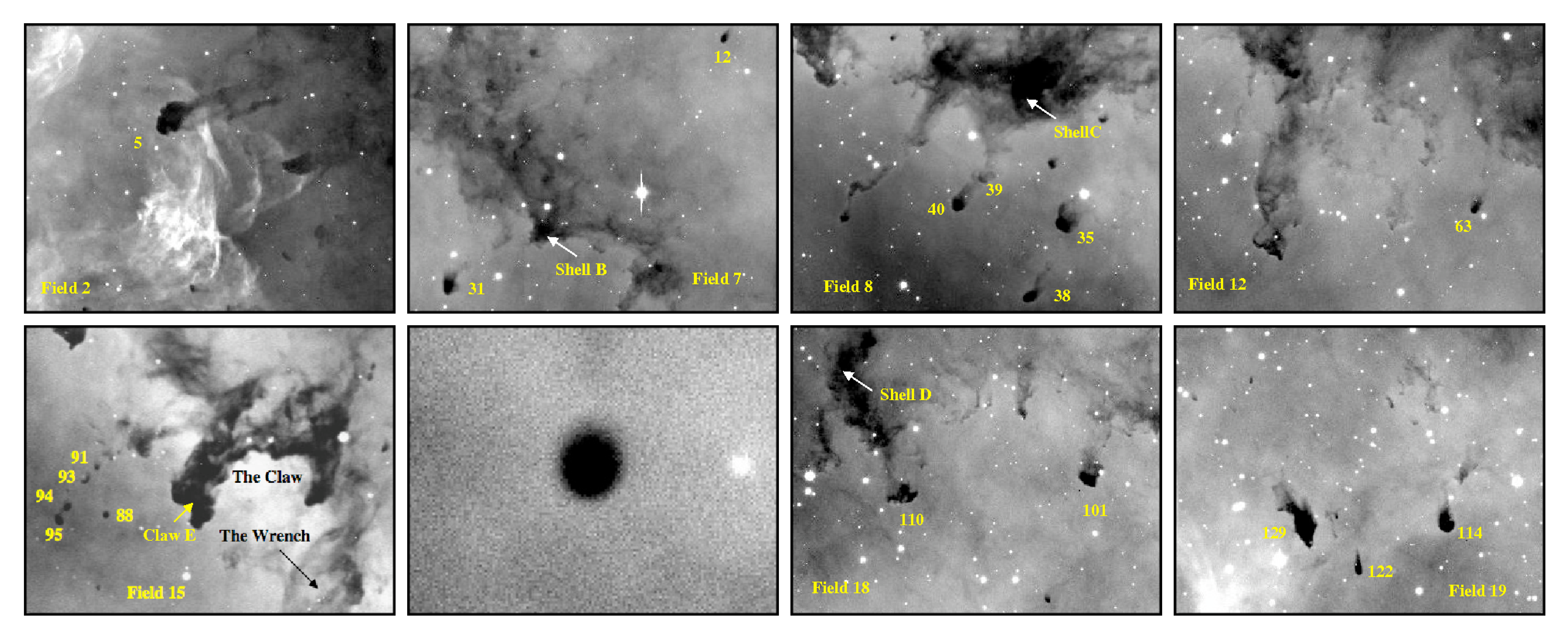}
\caption{Cuts from H$\alpha$ images presented in Paper~I  of different fields in the Rosette Nebula, including globulettes and positions in shells that were selected for the subsequent analysis, as numbered in 
Table~\ref{table:objects}. The fields all span 4.4$\arcmin$ x 3.4$\arcmin$, and are numbered as in Paper~I. The bottom second panel is a blown-up H$\alpha$  image of the dense globulette RN~88, diameter $\sim$~6~$\arcsec$ in Field~15. The string of globulettes in Field~15 is connected to a long filamentary shell containing a claw-like feature. The upper part of the prominent elephant trunk  "the Wrench" is seen southwest of "the Claw". North is up and east is to the left.}
\label{fig:fields}
\end{figure*}

Another four globulettes located in areas outside the optical survey were observed at OSO; one of them was included in 
the APEX observations. In addition, four positions in molecular shells in regions adjacent to globulettes
were observed at OSO. The NIR images contain several globulettes and trunks, and the results of
this survey will be discussed in more detail in a follow-up paper. Ten of the radio- detected globulettes fall in the NIR fields -- these data are considered in the present article as well.

The observed objects are listed in Table~\ref{table:objects} and ar identified according to their catalogue 
number in Table~2 of Paper~I. Here RN denotes the Rosette Nebula. Central positions of our radio observations 
were also taken from Paper~I. Then follows the objects added to this list, globulettes (marked A, B, etc.) for which central positions were selected from existing sky views of the Rosette Nebula, and shells (marked Shell~A, B, etc.) with positions measured from NOT images. The object designated {\it Claw~E} forms the eastern part of a yaw-like structure, composed of twisted and very thin filaments. The western part of the Claw was included in the CO survey by Gahm et~al. (\cite{gahm06}) of the adjacent "Wrench Trunk", which also has an yaw-like end (Carlqvist~et~al. \cite{carlqvist03}). 

{\it Columns 2} and {\it 3} in Table~\ref{table:objects} list sources observed for $^{12}$CO and $^{13}$CO $J$\,=\,1\,--\,0  at OSO. Objects mapped over a small area around the central position in selected lines are marked. The APEX observations marked in {\it Column 4} include all four observed lines. In {\it Column 5} objects that are included in the NIR fields are marked. 
In the the last column, objects also observed in HCO$^{+}$(1\,--\,0) are marked. RN~40 is identical to TDR~10 as designated in 
Gonz\'alez-Alfonso \& Cernicharo (\cite{gonzalez94}), and RN~A is identical in position to the highly obscured X-ray source ACIS~\#86 listed by Wang~et~al. (\cite{wang10}). Among the CO-detected clumps and distinct objects in Dent~et~al. 
(\cite{dent09}), no~8, here referred to as D~8, is close in position to this X-ray source. From the same list, D~10 is  identical to RN~5, and  D~6 includes objects RN~35 and 38, not resolved in Dent~et~al. (\cite{dent09}). D~15 is identical to 
RN~D (off-set by $\sim$7$\arcsec$).

\begin{figure}[t]
\centering
\includegraphics[angle=00, width=8cm]{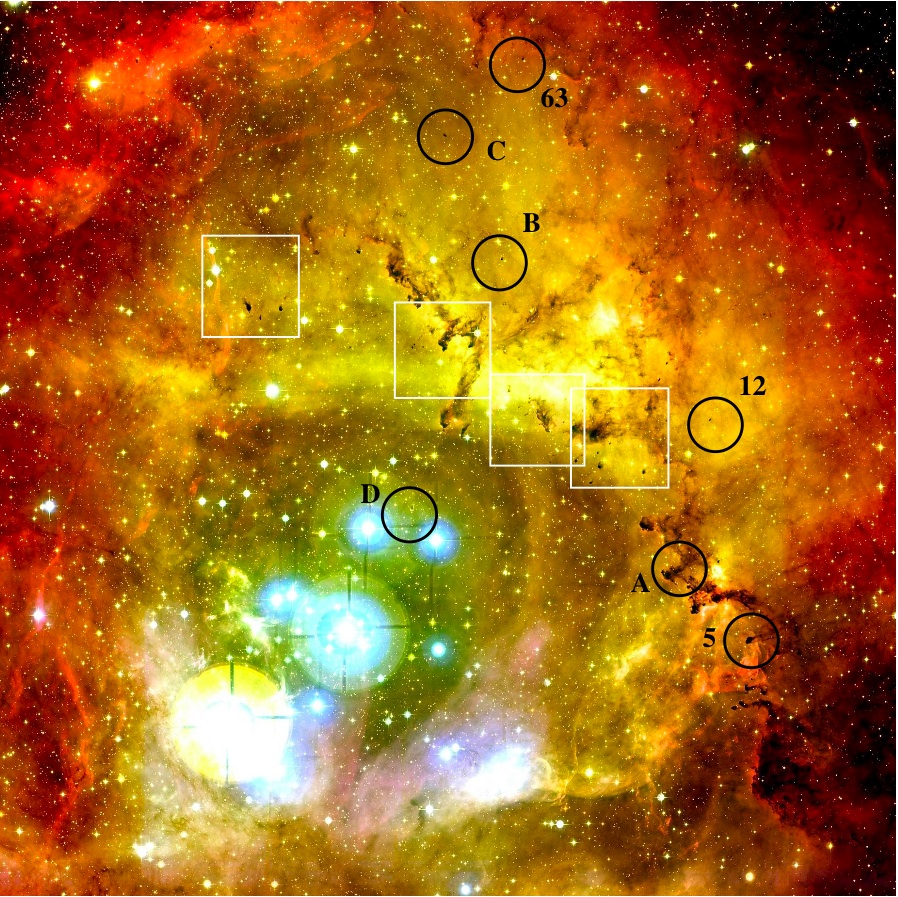}
\caption{Overview of part of the Rosette Nebula composed from images in H$\alpha$ (red) and \ion{O}{iii} (green). Objects not included in the NOT survey are marked, as are some NOT objects for orientation. The NIR fields are depicted as white boxes, covering from right to left Fields 7, 8, 15, and 19. The elephant trunk the Wrench is in the middle of the image with the Claw at its upper end (image: Canada-France-Hawaii Telescope).} 
\label{fig:view}
\end{figure}

\begin{figure}[t]
\centering
\includegraphics[angle=00, width=8cm]{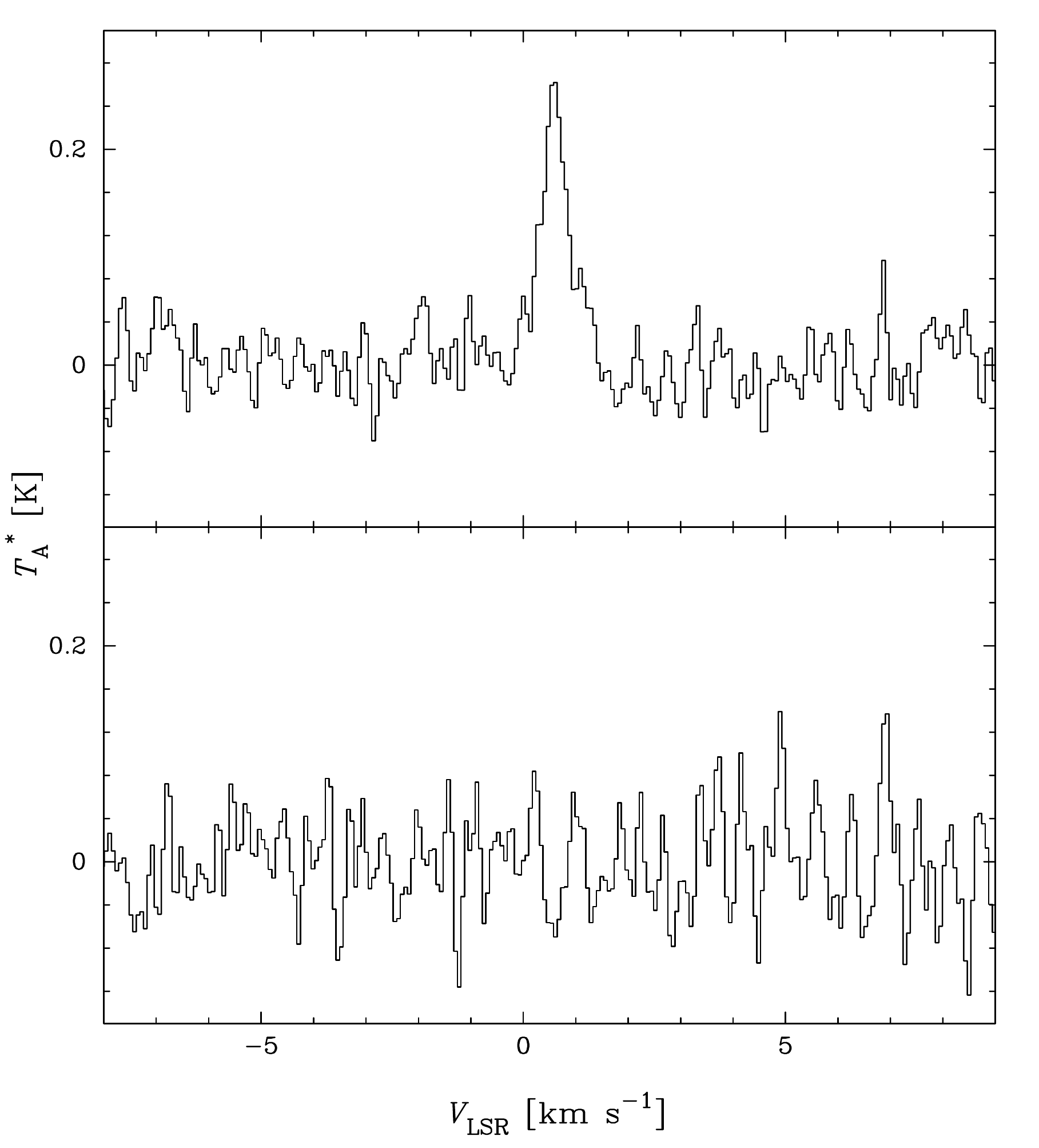}
\caption{$^{13}$CO(1--0) emission  towards RN~40 over the LSR velocity range -8 to +9~km~s$^{-1}$. The top panel shows a 
10-sigma detection at the on-position, and the lower panel shows the non-detection at the off-position (30$\arcsec$, $-30\arcsec$) with an rms of 45~mK.}
\label{fig:onRNoff}
\end{figure}

Figure~\ref{fig:fields} shows H$\alpha$ images, that were extracted from the fields in the 
NOT survey and include a number of globulettes listed in Paper~I. The locations of the fields in the Rosette Nebula can be found in Fig.~1 of Paper~I from the Field number. Objects observed in the present investigation are marked according to the designations in Table~\ref{table:objects}. Also shown is an enlargement of RN~88 in Field~15  as an example of a well-confined and dense isolated globulette. RN~88 appears to be completely dark in this high contrast image, but weak residual H$\alpha$ emission covers the object. In Field~8, RN~39 is marked because it is included in the CO map of TDR 10 in Gonz\'alez-Alfonso \& Cernicharo (\cite{gonzalez94}), where RN~40, separated by 24.5$\arcsec$, is the stronger source. However, RN~39 is outside our telescope beams. 

Some globulettes are connected by very thin filaments to shell structures, as in Field~18, and others are quite isolated from trunks and shells, as in Field~19. Both round and elongated globulettes are represented. In Field~15 a remarkable string of globulettes extends to the east of the Claw. 

The object RN~A, centred at an X-ray source, is not a regular globulette. It is attached to the shell and has the form of a block with finger-like extensions. Our radio and NIR observations reveal that this region is highly complex and contains an outflow; this object will be analysed in more detail in a separate paper. RN~A is therefore not included in the subsequent analysis of globulettes in the Rosette Nebula and is not displayed in Figure~\ref{fig:fields}.  RN~B and C are two distinct globulettes located in areas outside the NOT survey north of the inner molecular shell. RN~D is seen projected on the central cavity not far from the central cluster. The locations of all these objects can be found in Fig.~\ref{fig:view}, where some NOT-included objects shown in Fig.~\ref{fig:fields} are marked. The Wrench is in the middle of the image with the yaw-like feature the Claw. The position of Shell~A falls inside the RN~A circle. The four NIR fields are also marked. Their central positions are 06:31:00.9 +05:07:03.6, 06:31:17.9 +05:08:05.0, 06:31:39.8 +05:11:16.1, and 06:32:20.8 +05:14:06.0, and cover areas in fields 7, 8, 15, and 19.

\section{Results}
\label{sec:results}

\subsection{Spectral line parameters}
\label{sec:spectra} 

Molecular line emission was deteceted from all objects, with the exception of the small globulettes RN~12 and C. All our APEX and Onsala spectra at on-source-positions of objects detected in a least one transition are shown in the on-line Figs.~\ref{fig:AppendixA1} and \ref{fig:AppendixA2}. We are confident that we can distinguish the targets from other sources, since no signal is detected at off-positions. An example is shown in Fig.~\ref{fig:onRNoff} with \mbox{$^{13}$CO(1--0)} spectra obtained for RN~40 at on- and off-position (0$\arcsec$, -30$\arcsec$). The intensity scale in all spectra is expressed in terms of the antenna temperature reduced to outside the atmosphere, $T_\mathrm{A}^*$. Most objects show distinct and narrow lines, $\sim$ 1.0 km\,s$^{-1}$. The line peak velocities agree well for lines from various transitions with some exceptions. For several objects, emission from shells in the background enters, but with velocities different from the globulettes. At off-set positions, the globulette signal is gone while the background emission remains. 

\begin{table*}
\centering
\caption{Data and line parameters for observed globulettes. }
\begin{tabular} {llll | ll | ll|   ll | l }
 \hline\hline     
&&&\multicolumn{2}{c} {$^{12}$CO}&  \multicolumn{2}{c} {$^{13}$CO}    \\
RN & Size & Mass &Transition& $\int{T^{\,*}_\mathrm{A}  \mathrm{d} v}$\tablefootmark{a} & $\Delta v$\tablefootmark{b}&   $\int{T^{\,*}_\mathrm{A}  \mathrm{d} v}$\tablefootmark{a} & $\Delta v$\tablefootmark{b} & $v_\mathrm{LSR}$(B)\tablefootmark{c} & $v_\mathrm{LSR}$(G)\tablefootmark{d}& Remarks\\
& (arcsec$^{2}$) &     ($M_{J}$)     &       & (K km s$^{-1}$)  & (km s$^{-1}$) &   (K km s$^{-1}$)  & (km s$^{-1}$)  & & &\\
     \noalign{\smallskip}
     \hline
 5   &  384   & 665  & 1--0     & 4.5        & 2.6     & 0.7    & 1.0     & + 18.1 & + 1.9 & com., HCO$^+$  (0.3) \\  
   &          &     & 2--1     & 8.8        & 1.2       & 4.1    & 0.9       &              &   &  \\  
   &           &    & 3--2      & 8.0       & 1.2        &   3.7     & 1.0  	   &   &  &  \\
9 & 52    & 72   & 1--0     &  0.3           & 0.8                & 0.03 & 0.4                   & +2.2  &   - 0.5 & \\ 
   &          &     & 2--1     & 1.7           & 0.9         & 0.4        &   0.6    &   &     & 	  \\  
   &    &           & 3--2      & 1.6          & 0.9    & 0.5        &  0.6    	 &             &	&    \\
12&  41 &   47   & 1--0     &                 &           &     (*)     &         &    +17.1	&     &  \\  
31&  105 & 155  &  1--0    &  0.8          & 1.4     & 0.1       &   1.6    &	               &    -1.1  &  \\
35&     101 &  172  &  1--0    &          &            & 0.1       &   0.7     &	     &   -0.9  &  \\
38&    86  &  97   &  1--0    &  0.7          & 1.2      &  0.09       &   0.6       &   +12.7, +18.0 &    -0.4  &  \\
   &   &            & 2--1     & 3.1           & 1.2        &  0.7       & 0.7       & 	 & \\
   &     &          & 3--2      & 2.4          & 1.1    & 0.7        &  0.7         &	 &   &\\  
40&   102 & 123 & 1--0     & 1.1    &  1.4       & 0.2       &   0.7       & +17.1  &  +0.6  &  HCO$^+$ (0.1)  \\
   &   &            & 2--1     & 3.9           & 1.1       &  0.9       & 0.7       & 		&     &	  \\
   &    &          & 3--2      & 2.4          & 1.1   &  0.9        &  0.7         &		 &     & \\  
63&    54 &  59    &  1--0    &  0.4         & 0.7      &       (*)           &        & +17.7   &    +3.4  &  \\
 88&    21&  20   &  1--0    &                &              &       0.08  & 1.0         &               &  -0.3   &    \\
   &     &          & 2--1     & 0.5           & 0.8         &  0.2       & 0.6    & 	  -1.6	      &       & \\
   &       &        & 3--2      & 0.6          & 0.9      & 0.2        & 0.5         & &\\  
93/91& 44/17 & $>$14/12 &     1--0         &   &       & 0.09       &   0.7     &	  -0.2	 &   +1.2  & \\
 &     &          & 2--1     & broad                  & broad       &  broad       & broad        & 	 &    &\\
 &      &         & 3--2      & broad                 & broad   &  broad      &  broad & & &\\  
95/94&  34/21& 37/22 & 1--0    &  0.7          & 2.2     & 0.08       &   0.7     &    -0.4 &   +1.2  &   \\
   &   &            & 2--1     & 2.2             & 1.1       &  0.7       & 0.9          &  &   & \\
   &   &            & 3--2      & 1.9          & 1.1    & 0.6        &  0.8         &	&   &\\  
101&   108 &   234 &   1--0        &   &         & 0.1       &   0.9           &	  &   -2.4  &  \\
110 &  169  & 190  &   1--0  &    &     & 0.03       &   0.7      &  &    -4.0  &  \\
114&   145 &  195   &      1--0       &        &   & 0.2       &   0.7      &    &    -0.2  &  \\
   &    &           & 2--1     & 4.4           & 1.1    &  1.1       & 0.7       &   &    & \\
    &     &          & 3--2      & 4.2          & 1.2  & 1.1        &  0.7     &	&   & \\  
122&   50  & 68  &  1--0    &  0.5          & 1.1  & 0.09       &   1.0     &         &   +2.3  &   \\
  &      &         & 2--1     & 1.6           & 0.9    &  0.4       & 0.6       &   	& \\
   &      &         & 3--2      & 1.8          & 1.0   & 0.5        &  0.7     &		& & \\  
129&  364 & 691 &  1--0  &  1.9  & 1.6   & 0.4  & 1.0 &    & +1.8  & com., HCO$^{+}$  (0.3)    \\
   &   &            & 2--1     & 7.9           & 1.5    &  2.4       & 0.9       &   &    &    \\
   &    &           & 3--2      & 6.4          & 1.4  &  2.6        &  1.0     &	&   &    \\  
A&      &          &  1--0    &   5.2           & 2.0          &   &   & +8.4 &    +2.0  & com.   \\
   &     &          & 2--1     & 22.0           & 2.1    &  8.3     & 1.5      & &   &\\
   &    &           & 3--2     & 15.1           & 2.1   &   6.0    & 1.5       & &   &\\
B&      &         &  1--0    &   0.4           & 1.2          & 	&  &  &       -0.3  &   \\
C&     &         &  1--0   & (*)    &            & (*)  &   & +17.0 &   &        \\
D&     &          & 1--0      &   &   & 0.2       &   1.1     &       &  +2.8  &  \\
Shell A&   &    & 1--0     &  8.9        & 2.7   &   &   &   &  +2.4      & \\
Shell B&   &   &  1--0     &  2.0     &1.8    &  &   &    &   -0.5     & \\
Shell C&   &   & 1--0      &  7.9       & 2.3  & 2.3 & 2.0    &   	&  +1.6      & \\
Claw E&  &   &  1--0     & 5.3   & $\sim$2.0 &      && 	&    +1.3    & com.\\
Shell D&   &   &  1--0     &   1.4      & 2.4     &  &&	&   -4.4     & \\

\hline
\label{Table: transitions}
\end{tabular}
\tablefoot{
\tablefoottext{a}{Integrated intensity.}
\tablefoottext{b}{Line width from a Gaussian fit.}
\tablefoottext{c}{LSR velocity of the background emission.}
\tablefoottext{d}{LSR velocity of the globulette.}
\tablefoottext{*}{not detected.}
}
\label{table:spectra} 
\end{table*} 

In Table~\ref{table:spectra} (also available from CDS) we present line parameters derived for all objects and transitions. Objects with broad lines are so marked. The globulettes are normally well confined in the H$\alpha$ images presented in Paper~I, and size in {\it Column~2} refers to the area measured by us from the optical images, not including faint obscuring tails and plumes. The optically derived mass from Paper~I expressed in Jupiter masses is listed in {\it Column~3}. In {\it Column~9} velocities of the most prominent background components are given, to be discussed in Sect.~\ref{sec:background}. These components are as a rule strong and broad and at velocities very different from the globulettes. {\it Column~10} gives the \mbox{$^{13}$CO(1--0)} velocity if observed, otherwise it is obtained from a $^{12}$CO line. In {\it Column~11} objects with complex profiles (com.) that may contain more than one component are noted (futher discussed in Sect.~\ref{sec:discussion}). Objects observed for HCO$^{+}$ are indicated with integrated intensity in parenthesis. 

\begin{figure*}[t]
\centering
\includegraphics[angle=00, width=15 cm]{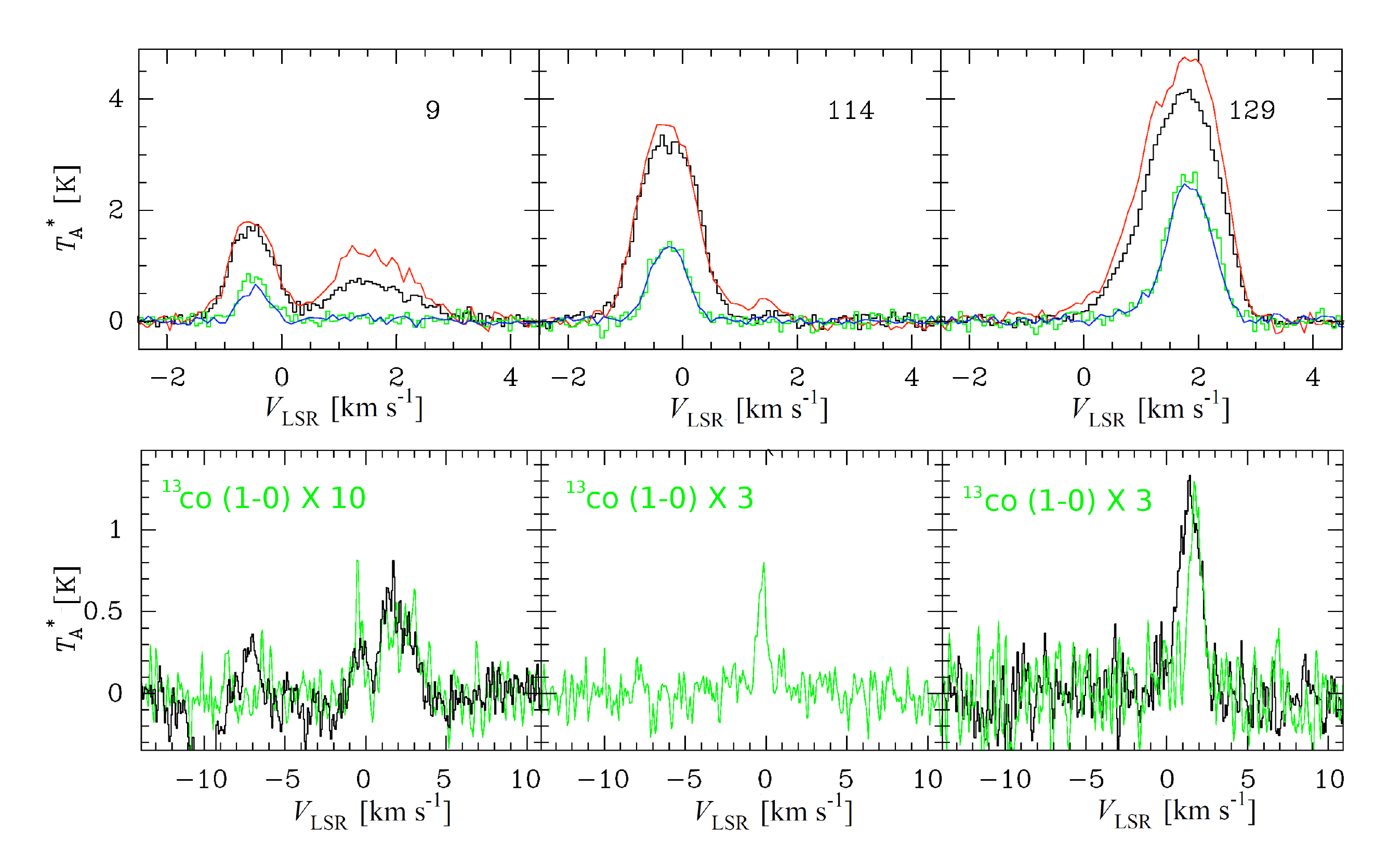}
\caption{Examples of spectra obtained with APEX (upper panels) and at OSO (lower panels) for globulettes of different size. From left to right: RN~9 (small), 114 (medium), and 129 (large). Upper panels: the  $^{12}$CO(3--2), $^{13}$CO(3--2), $^{12}$CO(2--1), and  $^{13}$CO(2--1) lines are plotted in black, green, red, and blue, respectively. Lines from (2--1) transitions are drawn as curves, and from (3--2) transitions as histograms. Lower panels: \mbox{$^{12}$CO(1--0)} profiles (thick); \mbox{$^{13}$CO(1--0)} profiles (green, thin). The  $^{13}$CO(1--0)  lines are multiplied by ten (RN~9) or three (RN~114 and 129). RN~9 is identified with the narrow component at $\sim$ 0 km\,s$^{-1}$, while the broader component at $\sim$ +2 km\,s$^{-1}$ comes from more extended gas. In RN~129 a "blue" wing is present in several lines. }
\label{fig:spectra}
\end{figure*} 

For the subsequent discussion we divided the globulettes into three size groups depending on area according to Table~\ref{table:spectra}: small ($<$~80~arcsec$^{2}$), medium (80 -- 200~arcsec$^{2}$), and large ($>$~200~arcsec$^{2}$). It should be kept in mind, though, that all selected globulettes are among the largest listed in Paper~1. Examples of spectra obtained at several frequencies of globulettes of different size are shown in  Fig.~\ref{fig:spectra}.  RN~9 (left panels) is seen in projection against local extended emission responsible for the broader component at $\sim$~+2 km\,s$^{-1}$. At on-position also the narrow component at $\sim$0 km\,s$^{-1}$ enters the spectra and is identified with the globulette. Some profiles obtained for RN~129 are asymmetric (right panels). Objects with pronounced line asymmetries are discussed in Sect.~\ref{sec:local}. 

A general feature is that the $T_\mathrm{A}^*$ peak intensity of the strongest component is the same, or nearly the same, in the \mbox{$^{12}$CO(3--2;2--1)} and the $^{13}$CO(3--2;2--1) transitions irrespective of globulette size. This is surprising considering that the beam-filling in the $^{12}$CO(3--2) transition is higher by a factor 2.25 than \mbox{(2--1)}. The only notable exception to this is RN~40, with a \mbox{$^{12}$CO(3--2)} intensity of only 60\% of that in $^{13}$CO(2--1). The \twco and $^{13}$CO(1--0) lines are usually very weak compared with lines from the higher levels. When wing emission or a second velocity component is detected, the observed \twco line ratios varies over the line (Fig.~\ref{fig:AppendixA1}). 

In the area containing the tiny globulettes RN~88, 91, 93, 94, and 95 two velocity components are detected.  For these globulettes there is some confusion regarding which signal comes from the globulettes, and some lines are broad and shallow. Our identifications are based on the appearance of one off-set spectrum from $^{12}$CO(1--0), which is common to all objects. This region is discussed in Sect.~\ref{sec:velocities}. 

\subsection{SOFI results}
\label{sec:sofi}

\begin{figure*}[t]
\centering
\includegraphics[angle=00, width=14cm]{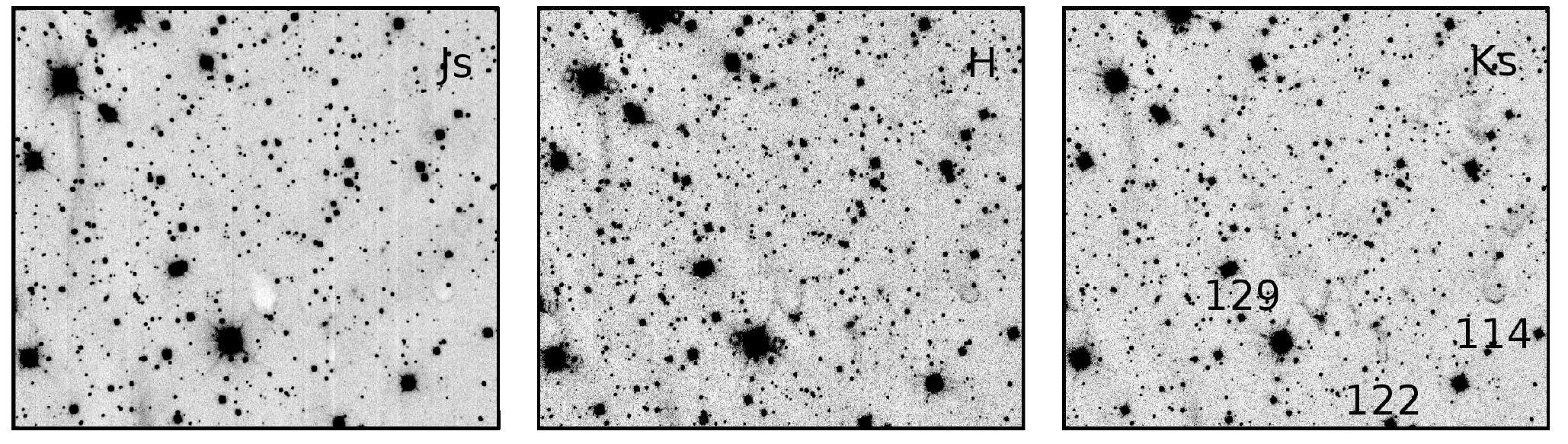}
\caption{Jittered Js, H, and Ks images of an area in Field 19. The cores of RN~114, 122 and 129 are opaque in the Js filter.}
\label{fig:jhk}
\end{figure*}

\begin{figure*}[t]
\centering
\includegraphics[angle=00, width=14cm]{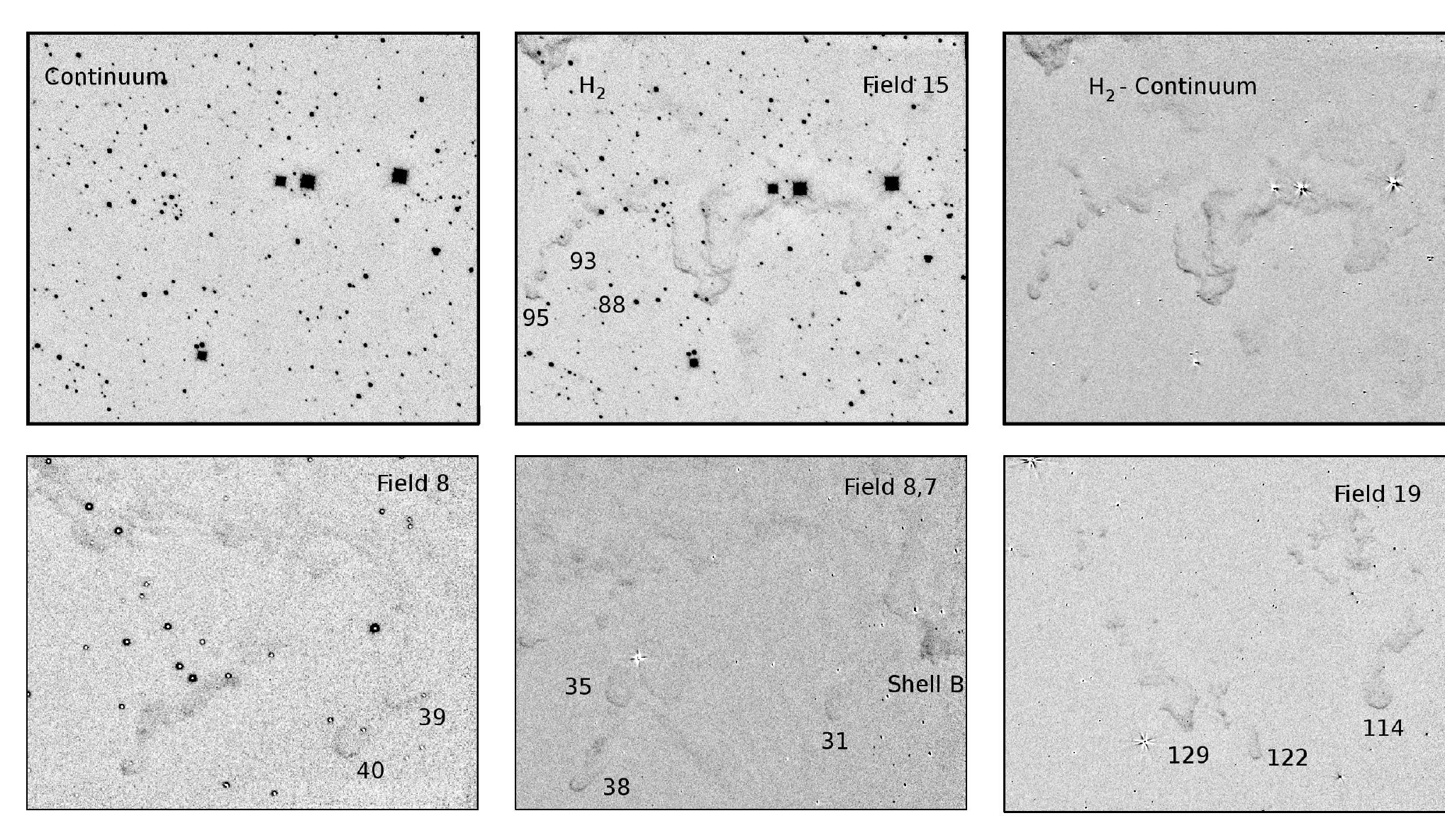}
\caption{Upper panels: continuum, \Htwo \ and the \Htwo\ -- continuum difference images of an area in Field 15 where the Claw and the string of small-size globulettes are visible in the two latter images. Lower panels: Difference images of areas in Fields 7, 8, and 15. The locations of some globulettes and one shell position are marked for orientation.}
\label{fig:H2}
\end{figure*}

\begin{figure*}[t]
\centering
\includegraphics[angle=00, width=14cm]{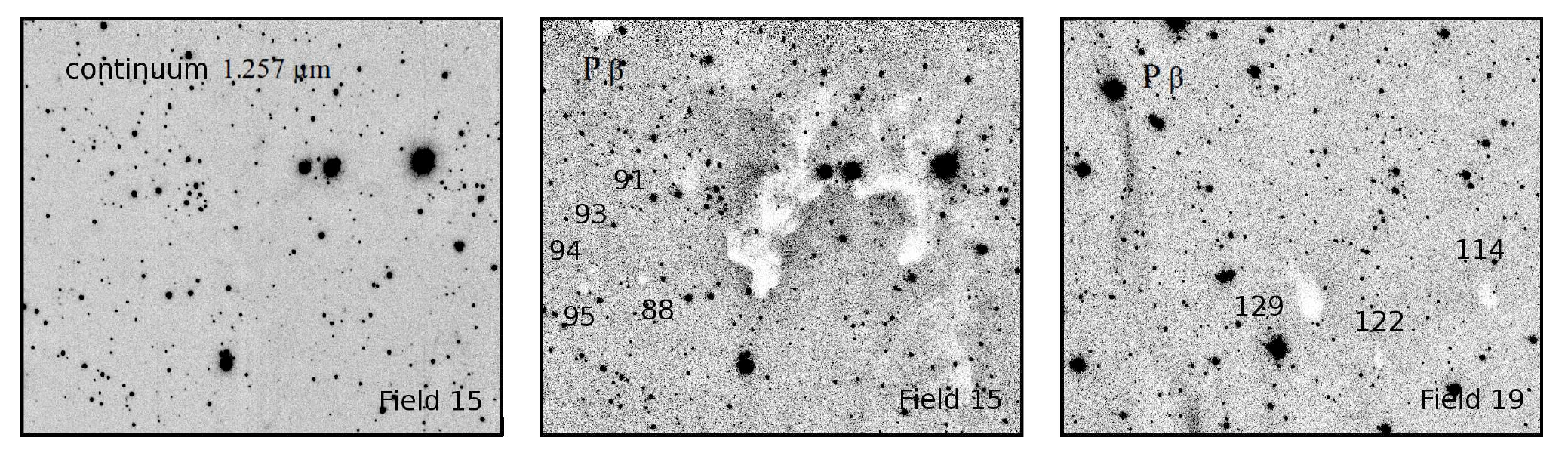}
\caption{Left and middle panels: 1.26 $\mu$m continuum and P$\beta$ images of an area in Field 15. The Claw at the centre is dark in the P$\beta$ image, and so are some of the tiny globulettes to the east of the Claw. Right panel: the opaque cores of RN~114, 122, and 129 are distinct on this P$\beta$ image of an area in Field 19 (see also Figure~\ref{fig:jhk}).}
\label{fig:Paschen}
\end{figure*}

The jittered \Js, \H\ and \Ks\ images of Field 19 are shown in Figure~\ref{fig:jhk}. Most of the globulettes, including the very small ones, are visible. In the \Ks\ images they appear as well-defined, faintly luminous bright-rimmed objects with sizes that match those seen in the NOT H$\alpha$ images. In the \Js\ image a few globulettes are seen as dark silhouettes, and several have bright rims as in the \Ks\ images.  The dark cores seen in the \Js\ images of RN~122 and 129 (left panel), for instance, are about half the size of those measured in H$\alpha$ and flag the presence of internal cores of higher density than traced in the optical survey. The jittering observing mode smooths out any extended surface brightness and furthermore, the data reduction forces the average background level to zero. Jittering may also smooth out to some extent the size and depth of areas of substantial extinction observed especially in the direction of the larger globulettes and shells and trunks.  Therefore we cannot derive NIR extinctions and infer column densities of dust, as was done in the optical spectral region (Paper~1). Some globulettes are also seen in the \H\ image, but they are less prominent than in the \Js\ or \Ks\ images.

The \Ks\ filter contains several molecular hydrogen lines and the hydrogen Brackett $\gamma$ line at 2.17$\mu$m. The observed fields were therefore observed also with the narrow-band filters NB H2 S1 covering the \Htwo\ 1-0 S(1) line at 2.12~$\mu$m, and NB 2.090, which covers the adjacent continuum. The difference between the two images contains only \Htwo\ emission, and all other features, including stars should vanish. The continuum, the \Htwo, and the difference images of Field 15 are shown in the upper panels of Figure~\ref{fig:H2}. The Claw and the string of small-size globulettes is seen in the \Htwo\ image, but not in the neighbouring continuum image. The difference image demonstrates that the globulette rims emit brightly in the 2.12 $\mu$m line. Faint \Htwo\ surface emission inside the globulettes is seen as well. NTT is an alt-azimutal telescope, and therefore the stellar diffraction spikes rotate on the images depending on the time of observation, and residuals resulting from the incomplete subtraction can be seen in the difference images. 

The lower panels in Figure~\ref{fig:H2} shows continuum-subtracted \Htwo\ 1-0 S(1) images in three areas located in Fields 7, 8, and 19. The incomplete elimination of stars in Field 8 was caused by the seeing, which varied between the continuum and \Htwo \ observations. The globulettes in these images have extremely thin bright rims, which are not resolved in the SOFI images. Bright rims also border trunks and other shell structures observed. A comparison of the \Htwo\ and the \Ks\ images shows that the \Htwo\ 2.12 $\mu$m line emission can explain approximately one third of the surface brightness observed in the \Ks\ image. Most probably, the major part of the excess emission is due to line emission in other \Htwo\ lines and the Brackett $\gamma$ line in the \Ks\ filter. Results from the \Htwo\ imaging will be discussed in detail more in a separate paper.

Bright rims, though fainter than in the \Ks\ images, are seen in the \J\ and \H\ images. These two filters also contain some rotational \Htwo \ lines and the hydrogen Paschen $\beta$ line at 1.28 $\mu$m enters the \J\ band. The 1.26 $\mu$m and P$\beta$ narrow-band filter images of the area selected in Field 15 are shown in Fig.~\ref{fig:Paschen} (left and middle panels). In the  P$\beta$ image the Claw and the string of tiny globulettes are seen dark against the bright background, whereas in the continuum image the objects cannot be detected. This is also the case for some objects in Field 19 (right panel).  

\subsection{SOFI stellar photometry} 
\label{sec:photometry}

\begin{figure}[t]
\centering
\includegraphics[angle=00, width=9cm]{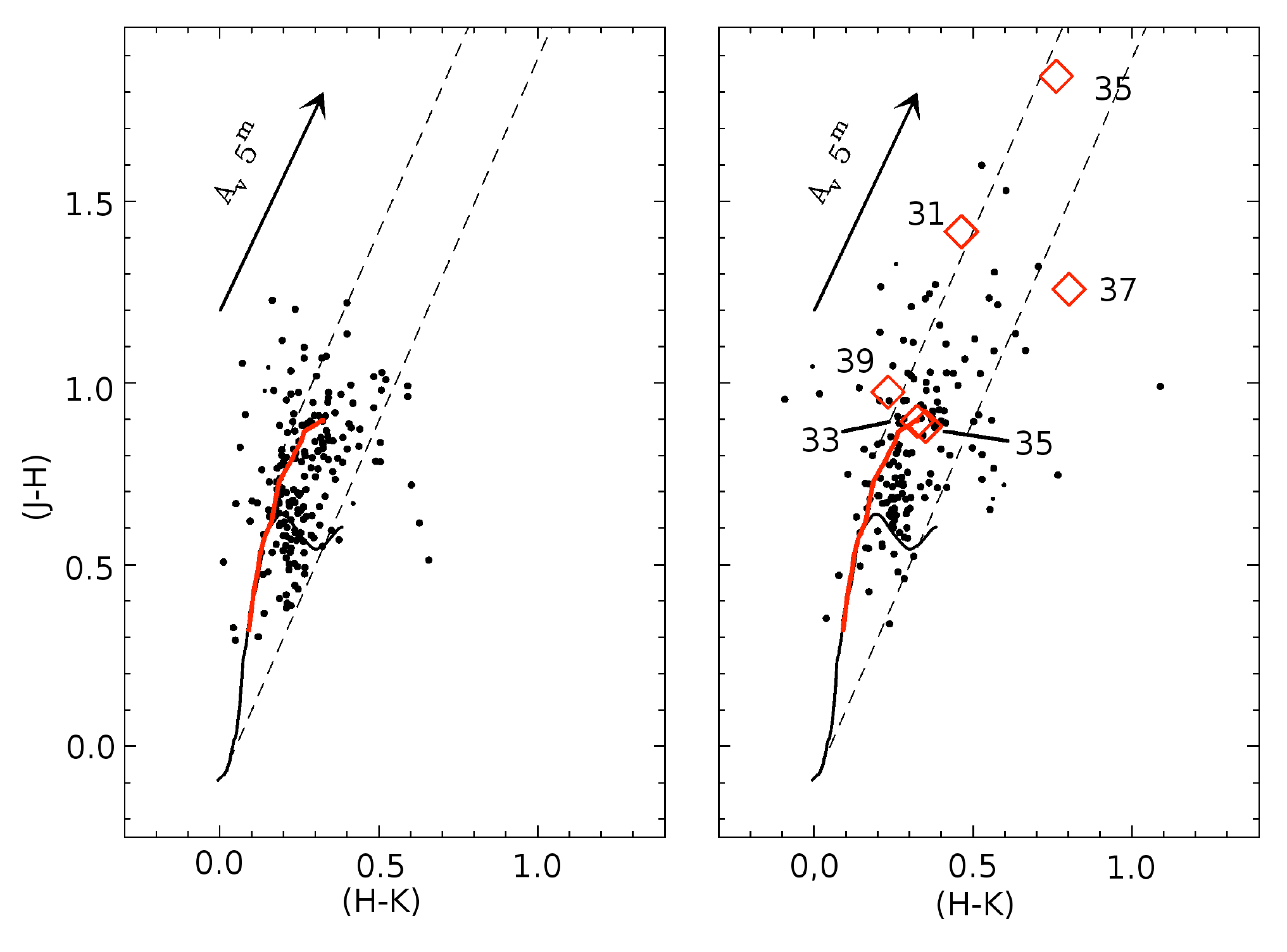}
\caption{$(J - H)/(H - Ks)$ colour-colour diagrams of stars from Fields 7 and 8. Stars outside the obscured region are plotted in the left panel, and those in the direction of shells and globulettes in the right panel. The stars detected inside globulettes are plotted as diamonds. Also plotted are the loci of unreddened giant (red, thick) and main-sequence (black, thin) stars. The direction of interstellar extinction follows the slashed lines and the vector shows a corresponding visual extinction of 5 mag. }
\label{fig:extinct}
\end{figure}

The stars in the SOFI field, which covers parts of the NOT fields 7 and 8 (Fig. \ref{fig:fields}), were divided into two to investigate the reddening in the direction of the general background and in the direction of globulettes and shell structures. The $(J - H)/(H - Ks)$ colour-colour diagrams of stars in these directions  are shown in Fig.~\ref{fig:extinct}.  The panel to the right includes stars in the direction of globulettes or shells, and the one to the left stars in surrounding areas. The arrows in the diagrams indicate the effect of five magnitudes of visual extinction. The left panel shows a grouping around the unreddened main-sequence, probably of unreddened or only slightly reddened late-type stars. The stars below the unreddened late main-sequence are possibly medium-mass stars. The uppermost grouping of stars in the left panel are stars in or behind the nebula. The location of these stars indicates a reddening of 1\fm5 to 2\fm0 of visual extinction.   

The number of substantially reddened stars is larger in the right panel. The grouping of stars around the unreddened main sequence is missing and only few stars fall below. The diamonds identify stars detected in the direction of globulettes in this field. Since these stars show no evidence of infrared excess emission, we conclude that they are reddened stars behind the globulettes, and not embedded very young objects.  From our data we cannot conclude whether the background stars are dwarfs or giants, but assuming a normal interstellar reddening law, the total extinction towards these globulettes is in the range $3 < A_{V} < 10$ magnitudes.

\section{Discussion}
\label{sec:discussion}

Using the information derived in  Sect.~\ref{sec:results}, we now proceed with a discussion of the physical properties of the globulettes and  how their velocities are distributed across the nebula. Because most of the globulettes are smaller than the HPBW of APEX at all observed frequencies, all spectral line modelling must rely on the physical sizes derived from optical and NIR images. The various images tracing both continuum and line emission allow us to draw some conclusions on density and density structure. Similarly to Paper~I, we adopt a distance of 1.40 kpc to the complex and a central velocity of +17~ km\,s$^{-1}$ (LSR) for the cluster and surrounding ionized gas in deriving physical quantities and estimates of mass. 

The globulette leading edges, i.e. the sides facing the central cluster, are sharp both in the optical and NIR. Some of the larger globulettes have shining bright edges in the H$\alpha$ line. However, if the background is bright, such rims may not be intense enough to be noticed. The southern edge of the Wrench and the edges of RN~73, 78--80, and 84 in Field 14 (see Paper~I) are bright, but those of the globulettes in Field 15 (RN~88, 94 and 95) and Field 19 are not (see Fig.~\ref{fig:fields}). 

Globulettes with bright H$\alpha$ rims also have bright rims in the \J\ and \Ks\ filters. The lack of bright rims in the narrow-band continuum images at 1.26 $\mu$m (Fig.~\ref{fig:Paschen}) and 2.09 $\mu$m (Fig.~\ref{fig:H2}) shows that the bright rims in the \Js \ and \Ks \ filters (Fig.~\ref{fig:jhk}) are not due to scattering but to line emission, as confirmed by the \Htwo\ and P$\beta$ images (Figs.~\ref{fig:H2} and~\ref{fig:Paschen}). The background in the P$\beta$ image is high but the Claw shows a distinct limb brightening. The globulettes just east of the Claw are detected in the image, but only in absorption against the background. Either bright rims do not exist, or they are very thin or faint and escape detection.

The bright rims observed in H$\alpha$ show that the density just above the globulette surface must be lower than that in the globulettes. Otherwise the optical depth at wavelengths below 1102 \AA, where the \Htwo\ absorption takes place, would be high and the radiation would not penetrate deep enough to reach the surface. Interstellar extinction at 1.282 $\mu$m (P$\beta$ line) is significantly lower than at 6557 \AA\ (H$\alpha$). Furthermore, the \Htwo\ rims coincide with the H$\alpha$ and P$\beta$ rims, which shows that matter turns into molecular form close below the surface. As stated in Paper~I, from the analysis in Grenman (\cite{grenman06}) the distribution of material in globulettes is not Gaussian, and the density is high even close to the globulette surface and is very low outside. Another sign of high density is that especially the larger objects are seen as dark silhouettes against the P$\beta$ background, which also applies to some of the smaller globulettes, such as RN~88, 91, 93, 94, and 95. We conclude that the globulettes are relatively dense and sharply bound objects consisting of molecular gas floating in a plasma of low density.  

\subsection{\Htwo\ column density and mass}
\label{sec:nir_mass}

Hydrogen is predominantly found in molecular form in the globulettes, therefore visual extinction scales directly with \Htwo \ gas column density, which can be estimated from the extinction to background stars in directions of globulettes. The NICER method presented in Lombardi \& Alves (\cite{lombardi01}) and the SOFI NIR photometry can be used to estimate the large-scale visual extinction within areas imaged in $\J\H\Ks$. In Sect. \ref{sec:photometry} we found that the extinction in areas outside globulettes, elephant trunks, and shell features is low, generally lower than 2 magnitudes. The extinction towards individual globulettes must be estimated individually, however. 

A highly reddened star is seen in the direction of globulette RN~35 (Fig.~\ref{fig:extinct}). Since the spectral type is unknown, one can only estimate upper and lower limits of the extinction. A star at the uppermost tip of the giant branch in Fig.~\ref{fig:extinct} would have a total visual extinction of  $\sim$ 7\fm5. The corresponding value for a star at the late unreddened main sequence or mid-giant branch is 10\fm0. 

The interstellar extinction to the Rosette complex is A$_{V}$~$\approx$~1.4 (Ogura \& Ishida \cite{ogura81}; Celnik \cite{celnik86}).  Allowing for 2\fm5 of foreground and also background extinction, the lower and upper estimates for RN~35 are 5 and 7 mag., respectively. If the star is even earlier in spectral type, the upper limit could be higher. Using the ratio of visual extinction to hydrogen column density given by Bohlin et al. (\cite{bohlin78}), the  extinction of  5\fm0  translates into an \Htwo\ column density of 4.7$\times10^{21}$~cm$^{-2}$. RN 35 is very symmetric in shape, and with a diameter of 18.6 kAU (Paper~I) this column density corresponds to an average density of 1.7$\times10^{4}$~cm$^{-3}$. With a mean molecular mass per \Htwo\ molecule of 2.8~amu, the total mass can be estimated to be 0.44\,M$_{\sun}$. This mass is about 2.5 times larger than the optically derived mass in Paper I, and if the extinction amounts to 7\fm5, the difference is even larger. The extinction derived for the star behind RN 35 probes only one direction through the cloud, however, and therefore the masses derived from the optical and NIR images are not directly comparable. 

\subsection{Molecular line observations}
\label{sec:molecularlines}

The interpretation of the molecular line data is not straightforward because we have no a priori knowledge of the small-scale globulette density, temperature, and velocity structure. The size of most of the observed globulettes is smaller than the APEX HPBW at 345~GHz and no mapping is possible. However, the high signal-to-noise ratio and good velocity resolution in the APEX spectra gives an opportunity to draw conclusions about the physical structure of globulettes. 

The solution of the radiative transport equation, neglecting background radiation and with a constant source function, is

\begin{equation}\label{solution}
T^*_\mathrm{A} =T_\mathrm{b}  \, \eta_{\mathrm{mb}}\, \eta_{\mathrm{bf}}=J(T_\mathrm{ex})  \, (1-e^{-\tau})  \, \eta_{\mathrm{mb}}\, \eta_{\mathrm{bf}},
\end{equation}
where $\eta_{\mathrm{mb}}$\ is the main beam efficiency, $\eta_{\mathrm{bf}}$ the beam-filling factor (see Eq. 3 and 4) and $J(T_\mathrm{ex})$ the radiation temperature

\begin{equation}
J(T_\mathrm{ex})  = \frac{h \nu}{k} \frac{1}{e^{\,h\nu/kT_\mathrm{ex}}-1} \approx T_\mathrm{ex},
\end{equation}
where the approximation is valid  if h$\nu$ $\ll$ k$T_{\mathrm{ex}}$.

A critical value in Eq. \ref{solution} is the beam-filling factor. The optical and NIR shape of the globulettes ranges from a nearly perfect, opaque  circle (RN~88, Fig.~\ref{fig:fields}) to cometary-like objects with a compact head and elongated, more diffuse tail (e.g. globulettes in Field 19, Fig.~\ref{fig:fields}). Besides the shape, the size also varies strongly. The cores of  interstellar dark cloud and normal globules are usually surrounded by less dense, extended envelopes, but as described above, the globulettes are more compact. The beam-filling of the compact round globulettes can thus be best estimated assuming a disk with a size corresponding to the projected optical size of the globulette. 

The beam-filling factor of a disk of diameter $\theta_\mathrm{D}$, a constant brightness temperature distribution over the disk and observed with a Gaussian beam with a half-width at half-maximum of $\theta_\mathrm{mb}$ is

\begin{equation}\label{eta_disk}
\eta_{\mathrm{bf}} = 1/(1-2^{-(D_\mathrm{DISK}/\theta_\mathrm{mb})2}).
\end{equation}

The beam-filling factors of symmetric, round globulettes can be estimated using the disk beam-filling factor. For elongated globulettes this is a combination of the disk filling factor (the leading edge) and the filling factor for a  Gaussian distribution (the tail). The beam-filling factor, $\eta_{\mathrm{bf}}$, assuming that both the source brightness distribution and the antenna response are circularly symmetric and Gaussian with a half width of  $\theta_\mathrm{s}$, is

\begin{equation}\label{eta_gauss}
\eta_{\mathrm{bf}} =\theta_\mathrm{s}^2 / (\theta_\mathrm{s}^2+\theta_\mathrm{mb}^2).
\end{equation}

The Onsala and APEX beams at the \twco frequencies (115, 230 and 345 GHz) are~33\arcsec, 27\arcsec and 18\arcsec, respectively. For a large-size source, filling the 345 GHz HPBW, $\eta_{\mathrm{bf}}$ is 0.19, 0.27 and 0.5 for a disk, and 0.23, 0.31 and 0.5  for a Gaussian distribution (Eq. 3 and 4).  For a small-size source like RN~88 (diameter 6\arcsec) the corresponding  values are 0.023, 0.034 and 0.075 for a disk, and  0.032, 0.047 and 0.1 for a Gaussian distribution. For globulettes with diameters of 10\arcsec\ or less the disk beam-filling factor is 25\% smaller than for the CO(3--2) transition, and more than 30\% smaller for the (2--1) transition.

The APEX main beam efficiency at 230 GHz and 345 GHz is nearly the same (0.75 and 0.73). Therefore, for a small size source with optically thick lines at the two frequencies. the observed $T^*_\mathrm{A}$, would depend strongly on the source-filling factor. Not surprisingly, the CO signal towards small globulettes is much weaker than towards the larger ones.

The $^{12}$CO(3--2)/$^{13}$CO(3--2) and $^{12}$CO(2--1)/$^{13}$CO(2--1) line ratios are approximately 1/3, irrespective of the source size. This indicates that both the $^{12}$CO(3--2) and (2--1) transitions are optically thick. The observed $^{12}$CO(3--2)/$^{12}$CO(2--1) and  $^{13}$CO(3--2)/$^{13}$CO(2--1) ratios do not depend on the source size but are close to one with the exception of RN~40, for which the \twco ratio is 0.6. 

The beam-filling factors described  above assume that the source brightness temperature has a Gaussian distribution or that the brightness temperature distribution is constant over the visible disk and that both the \twco  and \thco transitions probe the same volume of gas. However, it is unlikely that either assumption is valid and other factors have to be accounted for. In addition to the density, the CO line temperature depends on the excitation temperature. Assuming that the surface facing the central cluster is heated by radiation, a major part of the $^{12}$CO(3--2) emission could come from this warmer layer, while the (2--1) emission originates mainly from the cooler interior. The two transitions would then trace different volumes of gas, and the beam-filling factors would be different. Such a case was indeed observed by White et al. (\cite{white97}) in a cometary globule in the southern part of the Rosette Nebula. This type of source structure cannot be analysed using the LTE approximation, and a radiation transfer program is needed. 

\subsection{Globulette modelling}
\label{sec:model}

We have used a radiation transfer program developed by Juvela (\cite{juvela97}). The program uses the Monte Carlo method and allows sources with arbitrary density, temperature, velocity, density structure and a value for general turbulence to be constructed. Virtual observations of this model can be made in desired molecules/transitions. A different beam size can be chosen for each transition. The program calculates the appropriate beam-filling factor for the source structure and the beam in each transition, and thus the program output corresponds to the observed spectrum corrected for atmospheric attenuation and divided by the appropriate main beam efficiency. 

Test runs were made assuming a {\it standard} dense core structure with a cooler dense core region and a warmer, less dense envelope, where the density decreases towards the edges. The modelled \twco (3--2)/(2--1) line ratios were always much higher than the observed ones. The  radiation transfer program calculates the correct beam filling factor for each line, so the bad fit is not due to the better beam filling of the $^{12}$CO(3--2) line. Increasing the density and temperature of the envelope did not improve the situation. However, a better match to observed line intensities was obtained when a hot and thin dense surface layer was added to the latter models. Different versions of this toy model were tested.

Because only unresolved single-dish globulette spectra are available most of the input parameters must be guessed at. The observed spectra are narrow, less than 1~\kmps, which strongly restricts the value for the turbulence and the internal systematic (infall) velocity field. A turbulence of 0.2~\kmps~and an infall velocity of 0.1~\kmps~were chosen as an initial guess. Higher turbulence would produce too broad lines and higher in-fall velocity would, in addition to broadening the line, also produce spectra with a strong dip at the systemic velocity. For the smaller globulettes the values for the turbulence and in-fall had to be scaled down to obtain model spectra as narrow as observed. The density/temperature values were varied to obtain spectra similar to the observed temperatures.

To obtain the high densities needed to explain the observed NIR colours of background stars and to model the \thco\ lines, the \twco\ emission must be optically thick. The modelled \twco\ spectra are always strongly self-absorbed, so only lower limits for $T_\mathrm{ex}$ can be calculated from the observed $T_\mathrm{A}^*$. Taking into account the APEX beam-efficiency and the beam-filling factor as well as the observed $^{12}$CO(2--1) antenna temperature of 0.6~K for RN~88, the lower limit for this globulette would be well in excess of 15 K. For larger globulettes, such as RN~114, the lower limit would be 20 K. However, considering the likely self-absorption, the true $T_\mathrm{ex}$ must be higher. This agrees with the assumption that the globulette surface is heated by UV radiation from the central cluster. It is clear that a range of excitation temperatures must exist within each globulette.  

Even though it was possible to construct models that produced line intensities and line ratios similar to the observed \twco~and \thco~intensities and ratios, it was not possible to do this for both isotopologues simultaneously. A model that provides the correct \twco~intensities results in much too low \thco~intensities, and a model providing the correct \thco~intensities fails to explain the \twco~intensities. Two models were constructed for each globulette observed with APEX, one giving the \twco~line intensities and line ratios and another for the \thcop. The \Htwo\ densities required to model the observed \thco~spectra are significantly higher than those required for \twcop. The modelled \twco~spectra are strongly self-absorbed, which depends strongly on the source density and temperature structure and the velocity field (turbulence and systematic motions). 

We therefore used these models to provide a plausible lower (the \twco~model) and upper (the \thco~model) limit for the mass of each observed globulette. The adopted values for the velocity dispersion and systematic infall are probably adequate. The line width and shape of the spectra depend very strongly on these two parameters, and the pool of possible values is very restricted. No model that would produce the exact, observed  \twco~or \thco~line intensities could be constructed, and the $^{12}$CO(3--2) line was always 10 to 15\% stronger than the (2--1) transition as were also the corresponding values for the \thco~transitions. For the \twco~this may be explained by strong self-absorption. The upper and lower mass limits calculated from the models should therefore be considered to be only crude estimates.

As expected from the beam-filling argument alone, the larger the globulette, the stronger the observed molecular lines. The line width also increases with size. The \twco and \thco line widths increase from 0.9~\kmps\ and 0.6~\kmps\ to 1.3~\kmps\  and 0.9~\kmps, respectively. The morphology changes as well: the small-size globulettes are round and become more elongated with increasing size, i.e. tails emerge. The leading edge remains sharp, but the tail has a more diffuse appearance. If one assumes that the tail is formed by gas and dust flowing away from the head, then the outflowing gas should be blue-shifted with respect to the head because of the assumed orientation of the objects. This is indeed observed, and the more asymmetric the shape, the more pronounced the blue-shifted wing (see Sect.~\ref{sec:local}). 

{\it Small-size globulettes.} The physical diameters of  RN~9, 88, and 95 range from 8.5~kAU (RN~88) to 14~kAU (RN~9) (Paper 1, Table 2). The maximum \Htwo\ number densities used in the \thco~models are  around 1.1$\times$10$^4$~cm$^{-3}$ at the centre and 2.5$\times$10 $^{3}$~cm$^{-3}$ at the compressed edge.  All these globulettes are seen in absorption in the P$\beta$ image, which also supports the modelled high central density. Despite this high density the masses for the three objects are low. The mass lower limits from \twco\ models are 10 to 40 $M_{J}$ and upper limits from \thco\ models 40 to 70~$M_{J}$. The average optical masses (Table~\ref{table:spectra}) range from 20 (RN~88) and 72~$M_{J}$ (RN~9). Hence, the masses derived from visual extinction and radio line data are very similar. RN~95 has a blue-shifted velocity component, but this might be because the nearby globulette RN~94 also enters the Gaussian beams. Another two small globulettes, RN~93 and RN~122, were detected with APEX. The line profiles of RN~93 are very broad, however, and RN~122 is very elongated with a blue-shifted component probably related to its distinctive tail. Neither of these objects can be modelled with the present method. Their line intensities are similar to what is observed towards the modelled small-size globulettes, and presumably their masses fall in the same range.   

{\it Medium  size globulettes.} RN~38, 40, and 114 have diameters of 16, 17 and 22 kAU, respectively, and differ from the smaller globulettes in that they have distinctive tails in the H$\alpha$ images (Fig. \ref{fig:fields}). Only the interior cores, which are round, are opaque in the P$\beta$ images. The mass lower and upper limits derived using the spherically symmetric model are 70 and 130~$M_{J}$ for RN~38 and 100 and 345~$M_{J}$ for RN~114. Hence, the average masses for these objects also agree well with the values derived from optical data, namely 97 and 195 $M_{J}$. 

Modelling RN~40 is problematic because the $^{12}$CO(3--2) line intensity is only 60\% of the $^{12}$CO(2--1) intensity. The beam-filling factor of the 345~GHz beam for a Gaussian intensity distribution is 0.31, twice that for 230~GHz. It is impossible to construct a spherically symmetric model producing such a \mbox{$^{12}$CO(3--2)/(2--1)} ratio. In addition, blue-shifted wings are present (see Sect.~\ref{sec:local}).

RN~40 forms the head of TDR~10 observed in \twco and $^{13}$CO(2--1;1--0) and CS(3--2;2--1) by Gonz\'alez-Alfonso \& Cernicharo (\cite{gonzalez94}). TDR~10 also covers RN~39 situated 25\arcsec\ NW of RN~40. Curiously, their $^{12}$CO(2--1) line integral maximum is not observed in the head of RN~40, but in the tail just to the north. In the channel map (Fig. 3 in their paper) $^{12}$CO(2--1) is stronger at velocities $<$ 0.5~\kmps, i.e. on the blue-shifted side, compared with our $^{13}$CO(3--2), (2--1) and (1--0) central velocities. The $^{12}$CO(2--1) maximum would lie at the edge of the APEX 18\arcsec\ HPBW at 245~GHz, but still in the 27\arcsec\ APEX 230~GHz and the 33\arcsec\ Onsala 115 GHz beams. If the location of the $^{12}$CO(2--1) maximum in the tail were caused by a pointing error, correcting for such a  shift would move the secondary maximum, now coinciding with RN~39, to the south of it. Thus this pointing error is not likely.

The APEX observations were pointed at the RN~40 dense core, as evidenced by the \thco (3--2)/(2--1) ratio. Furthermore, the  velocities of all the observed \thco transitions agree. The larger beams at 230~GHz, and especially at 115~GHz trace the \twco emission maximum better than the 345~GHz beam. This brings us to the question: why the \twco maximum is seen in the RN~40 tail, which, at least in the optical H$\alpha$ image, appears diffuse and not opaque? One explanation could be that the \twco emission traces warm but diffuse subthermally exited gas.   
 
{\it Large-size globulettes.} Two large size-globulettes, RN~5 and 129 were observed with APEX. Both objects are elongated, more so than the APEX HPBW at 345 GHz, and have excess blue-shifted \twco emission. Curiously, in RN~5 the central line velocities of the $^{12}$CO(3--2) and (2--1) main line components are red-shifted compared with the \thco lines. Similarly to RN~40, the $^{12}$CO(1--0) line is broad and red-shifted compared with the higher transitions. It is possible to obtain a tentative mass estimate for the heads of the globulettes covered by the single-point observation, but not for the entire globulettes. No P$\beta$ image is available for RN~5, but in the H$\alpha$ image (Fig. \ref{fig:fields}) the head of the globulette, which possibly contains two fragments, is most opaque.  In the  RN~129 P$\beta$ image (Fig. \ref{fig:Paschen}) the beam covers the most obscured part of the globulette. The upper and lower limits for the estimated masses are 368 and 788 $M_{J}$ for RN~5, and 305 and 536 $M_{J}$ for RN~129. The \Htwo\ densities are similar to those needed to model the medium-sized globulettes above, but the masses are larger because of the size. Better beam-filling is the main reason for the higher observed antenna temperatures. The model cannot reproduce the strong blue-shifted emission, but it it assumed that this is due to blue-shifted gas in the more diffuse tail (Sect. \ref{sec:local}). The corresponding optical masses of 665 and 691 $M_{J}$ best match the maximum mass derived above, and were derived from slightly larger areas than covered by the APEX beams.    

\subsection{Global velocity pattern}
\label{sec:velocities}

The globulettes are distributed over a large area in the Rosette Nebula, spanning 1/2$\degr$ (a projected distance of 13 pc) from RN~5 to RN~63.  Most objects fall along the inner remnant shell, extending from west to northeast from the central cluster. As can be seen from Fig.~\ref{fig:view}, this shell forms part of a ring outside the northern part of the central cavity. The entire complex is blue-shifted relative to the centre of the Rosette complex by $\sim$~17~km\,s$^{-1}$. (Dent et~al. \cite{dent09}), and with an estimated inclination to the line-of-sight to the central cluster of $\sim$~40$\degr$~$\pm$~5$\degr$ (Schneps et~al. \cite{schneps80}; Gahm et~al. \cite{gahm06}).     

The velocity pattern is remarkably smooth. The globulettes all move with similar velocities, +0.5~$\pm$~2.1~km\,s$^{-1}$, very close to the velocities obtained at various positions along the ring. With an inclination of 40$\degr$ it follows that the system of globulettes, shells, and trunks expands at velocities of $\sim$~22~km\,s$^{-1}$ from the central cluster. The two objects farther out from the ring, RN~C and 63, have the highest positive velocities, +2.8~km\,s$^{-1}$ and +3.4~km\,s$^{-1}$, respectively. This is entirely consistent with an expansion velocity of 22~km\,s$^{-1}$, since in a spherically expanding system their inclination is $\approx$~60$\degr$. We did not consider here the geometry proposed by Dent et~al. (\cite{dent09}), who in an attempt also to include shell structures southeast of the cluster defined an expanding molecular ring centred far south of the cluster, and inclined by $\approx$~60$\degr$ (or $\approx$~30$\degr$ from the sky plane). This model implies a much higher expansion velocity of  $\sim$~30~km\,s$^{-1}$. For the northwestern area, at focus of the present investigation, we note that Viner et~al. (\cite{viner79}) obtained a line width of 22~km\,s$^{-1}$ from the H~110$\alpha$ recombination line, consistent with our result. For the same area Celnik (\cite{celnik85}) derived line widths of 15 and 31~km\,s$^{-1}$ from the He~112$\alpha$ and H~112$\alpha$ lines, respectively.    

RN~D, seen in projection to the cluster, also has a comparatively large positive velocity, +2.4~km\,s$^{-1}$, consistent with the value given in Dent et~al. (\cite{dent09}). In this direction one expects velocities of $\sim$ --3~km\,s$^{-1}$ for objects in the ring system. Hence, it appears that this isolated object seen against the central cavity has a different history than the ring system. 

The difference in radial velocity of globulettes and adjacent shell structures rarely exceeds 2~km\,s$^{-1}$. For example, RN~101 and 110 are moving with the most negative velocities in our sample, -2.4~km\,s$^{-1}$ and -4.0~km\,s$^{-1}$, but so does Shell D with -4.4~km\,s$^{-1}$. The globulettes are connected by thin filaments to Shell D. 

\begin{figure}[t]
\centering
\includegraphics[angle=00, width=8cm]{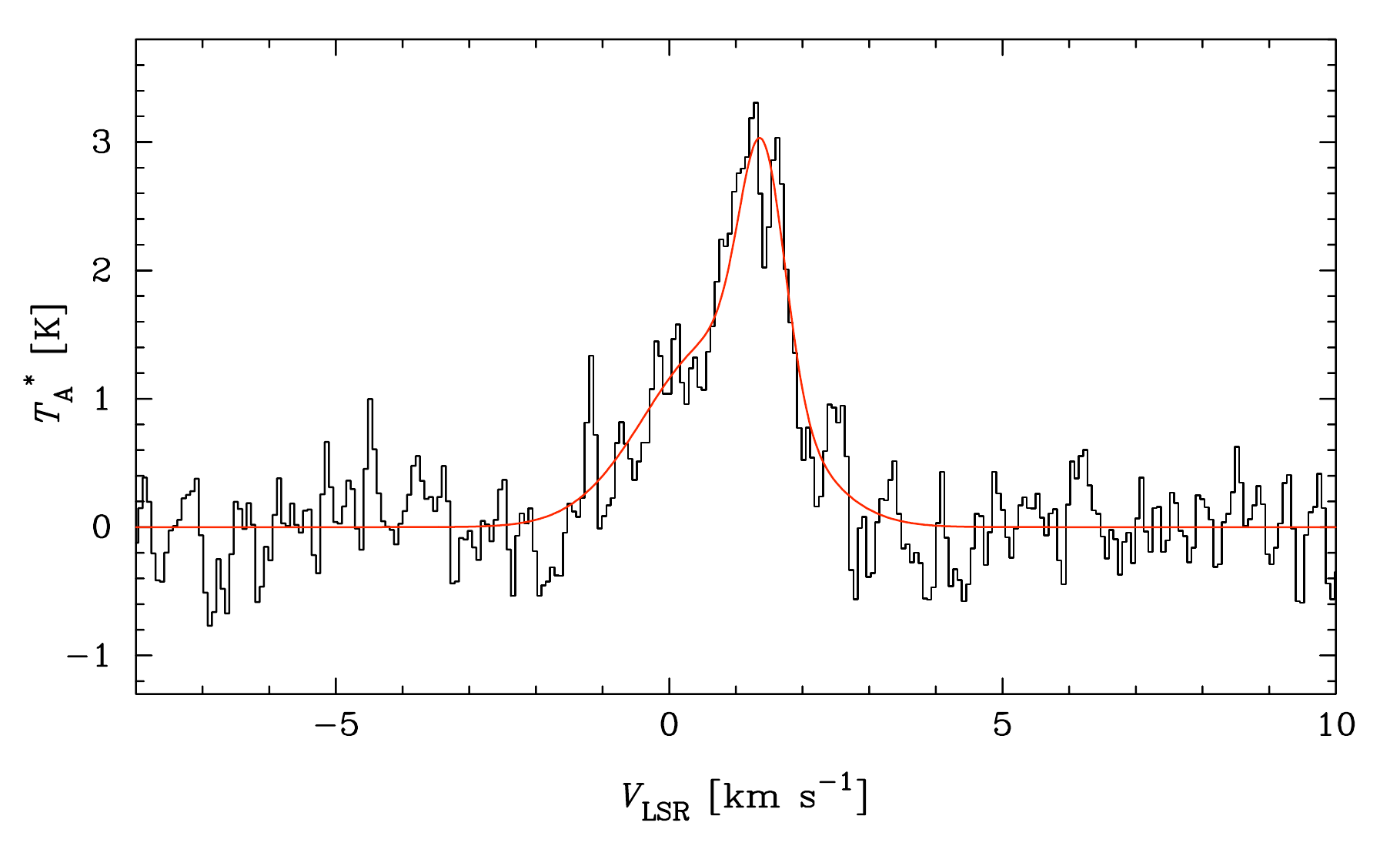}
\caption{$^{12}$CO(1--0) profile of the eastern part of the Claw, which is asymmetric,
but is as a whole red-shifted relative to the western part of the Claw. A two-component Gaussian fit to the profile is also shown (details in the text).}
\label{fig:claw}
\end{figure}

\begin{figure}[t]
\centering
\includegraphics[angle=00, width=8cm]{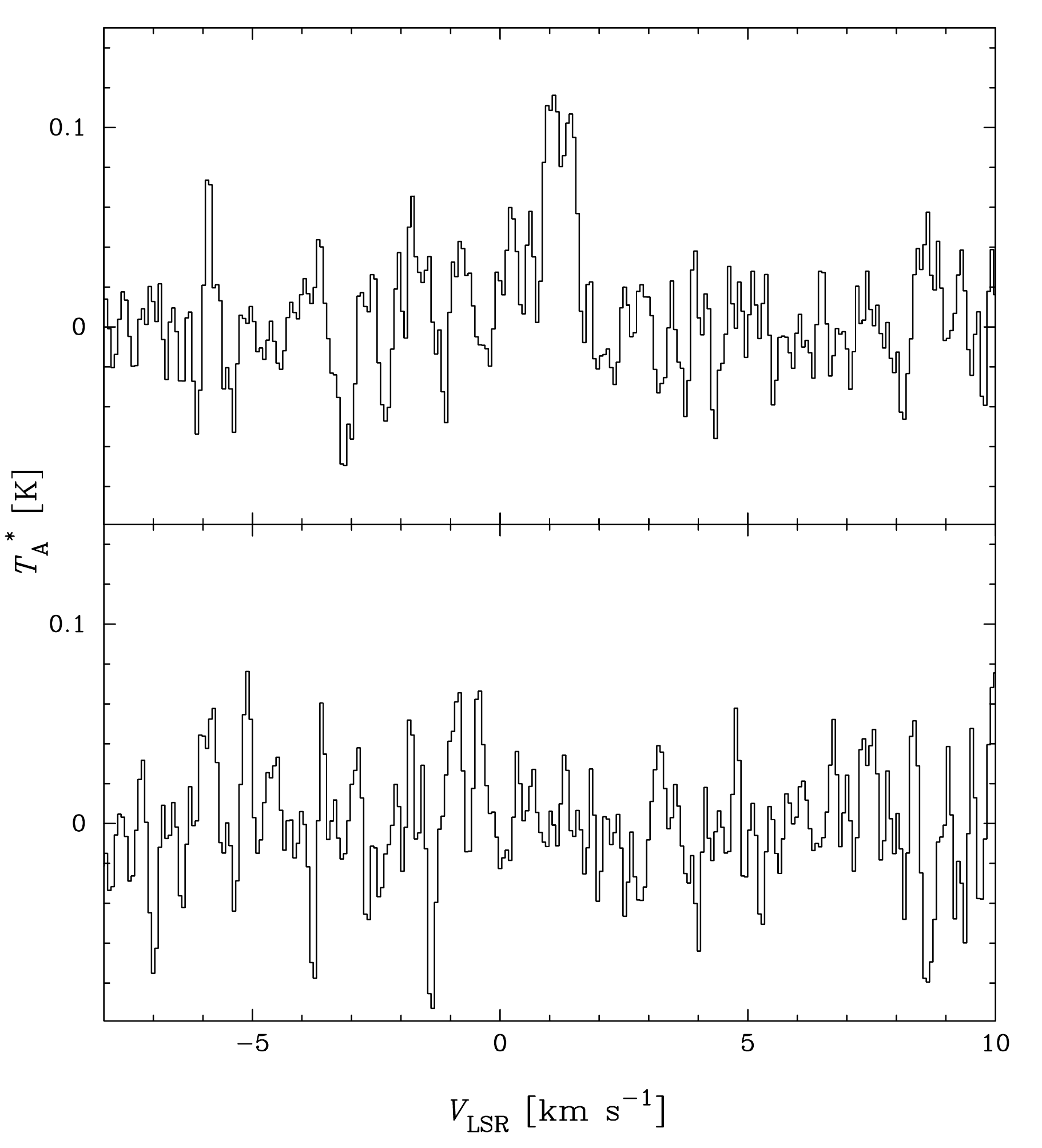}
\caption{$^{13}$CO(1--0) spectra obtained for the tiny globulette RN~95 at on-position (above) and off-position (+30$\arcsec$, 0$\arcsec$). }
\label{fig:RN95}
\end{figure}

Fig.~\ref{fig:claw} shows the profile obtained from the eastern part of the Claw (Claw E). A two-component Gaussian decomposition gives a narrow main component centred at +1.4~km\,s$^{-1}$ (FWHM =  0.8~km\,s$^{-1}$), and a weaker broader component at +0.7~km\,s$^{-1}$ (FWHM = 2.5~km\,s$^{-1}$). The western part of the Claw was included in the CO map of the Wrench by Gahm et~al. (\cite{gahm06}) and was found to move at $\approx$~+0.0~km\,s$^{-1}$. Hence, Claw E is red-shifted relative to the western part of the Claw by $\sim$~+1.2~km\,s$^{-1}$, which shows that the Claw as a whole rotates with about the same speed and the same direction as the Wrench, which is located just south of the Claw. 

RN~91, 93, 94, and 95 in the string east of the Claw (see Fig.~\ref{fig:fields}) have similar velocities, $\sim$~+1.2~km\,s$^{-1}$. RN~93 and 95 are larger than 91 and 94, respectively. Because  these pairs are separated by only $\approx$ 12$\arcsec$, the emission from the smaller globulettes has entered the radio beams, and there is also weak extended CO-emitting gas at similar velocities in this area. However, as demonstrated from the \mbox{$^{13}$CO(1--0)} spectra of RN~95 in Fig.~\ref{fig:RN95}, there is no problem to distinguish the globulette signal, which is detected in all transitions. The isolated object RN~88, just south of the string, is slightly blue-shifted in comparison, moving at -0.3~km\,s$^{-1}$. RN~88 is the smallest object in our survey, and the identification is tricky. However, the line ratios of the blue-shifted component are similar to the ratios observed in other globulettes, whereas the ratios in the red-shifted component are not.

\subsection{Local mass motions}
\label{sec:local}

We now take a closer look at three objects, RN~5, 40, and 129, which show more complex line profiles, an indication that additional local gas motions are present.

\begin{figure}[t]
\centering
\includegraphics[angle=00, width=8cm]{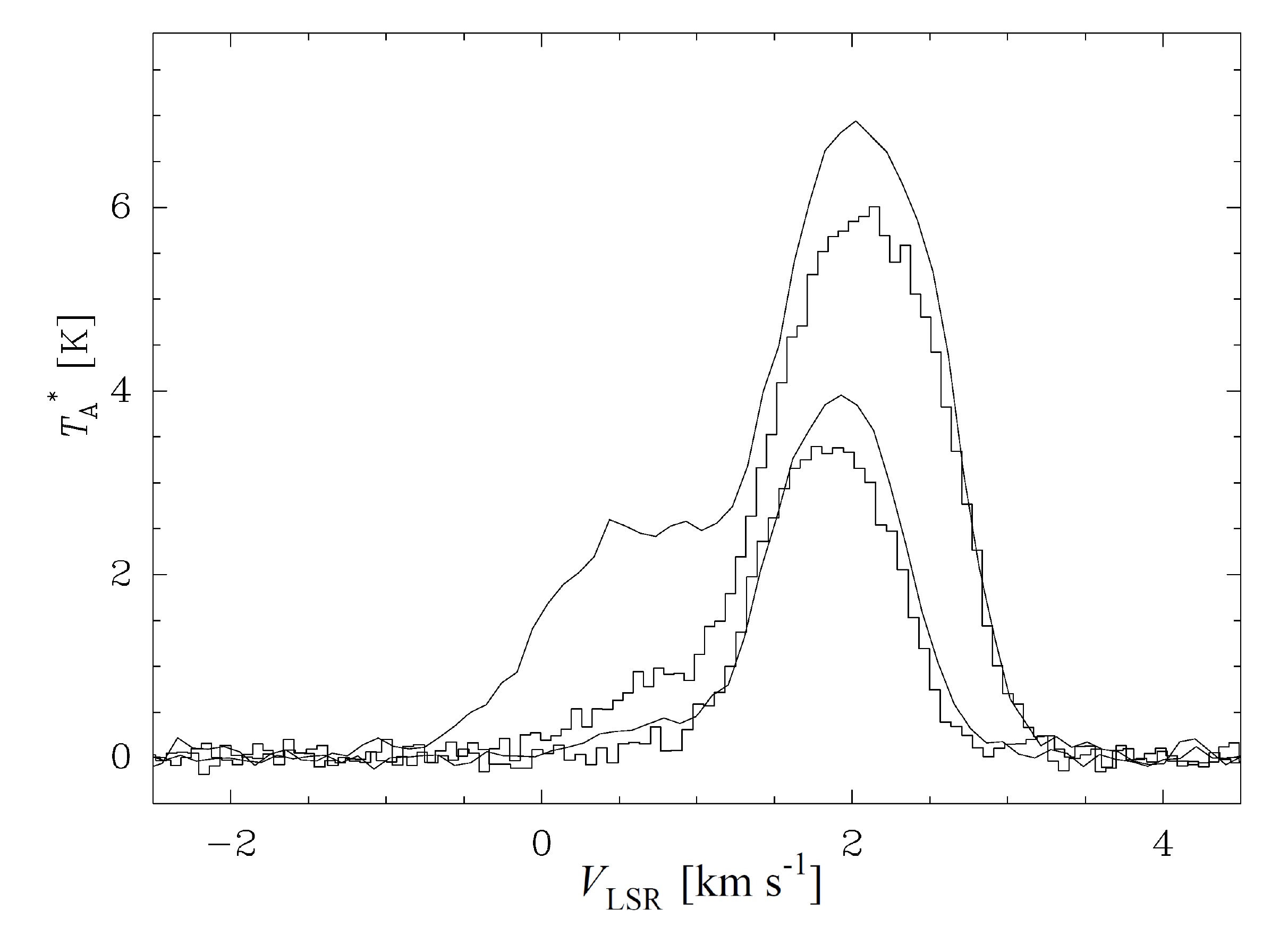}
\caption{Line profiles obtained for the central position of RN~5 show a weak component,  blue-shifted relative to the main component. The strongest lines are from $^{12}$CO and the weaker from $^{13}$CO. Lines from (2--1) transitions are drawn as curves, and from (3--2) transitions as histograms.}  
\label{fig:W5apex}
\end{figure}  

The most prominent wing feature is observed for RN 5, and in the APEX data the \mbox{$^{12}$CO(2--1)} transition is strongest and the \mbox{$^{13}$CO(3--2)}, weakest as seen in Fig.~\ref{fig:W5apex}. In RN~40 all \twco transitions have similar intensities in the wing, while in RN~129 the two higher \twco transitions are of nearly the same intensity, but the (1--0) intensity is also strong. A common feature in RN~5, 40, and 129 is that the \mbox{$^{12}$CO(1--0)} line is broader than that in the (3-2) and (2-1) transitions. The velocity of the (1--0) peak intensity is also shifted towards the wing.       

A two-component Gaussian fit to the Onsala profiles of RN~5 indicates that the weaker blue-shifted component moves with a velocity of $\sim$~1.4~km\,s$^{-1}$ relative to the main component. This object has a long and very pronounced tail extending to the west from the core (see Field 2, Fig.~\ref{fig:fields}); we identify the blue-shifted emission as coming from this tail. Since the main line component is comparatively broad, reaching FWHM = 2.6~km\,s$^{-1}$ in the \mbox{$^{12}$CO(2--1)} line, we mapped the region in \mbox{$^{12}$CO(1--0)} over a nine-grid matrix with steps of 10$\arcsec$ to search for any indication of outflows from an imbedded source. The CO-line in question can best be described by two components, but the line is optically very thick, and self-absorption may affect the line shapes. Our map does not reveal any systematic mass motions as expected from a bipolar flow, but the blue component is strongest on the western part of the globulette, supporting our suggestion that the blue-shifted emission arises in the tail.

The asymmetric line profiles of RN~129 are best seen in lines of \twco\ (Fig.~\ref{fig:spectra}), which are the most optically thick lines. RN~129 also has a pronounced tail, or rather plume, extending to the north and northeast of the core (see Field 19, Fig.~\ref{fig:fields}). As for RN~5, we identify the blue-shifted component with emission from this tail, which then would expand in our direction with a velocity of  $\sim$~0.6~km\,s$^{-1}$ relative to the core. RN~129 and RN~5 are the most massive globulettes in our sample, but RN~122, which is less massive, also has a distinct tail and a blue wing (Sect.~\ref{sec:model}). 

Finally, RN~40 has a blue-shifted wing present in the $^{12}$CO(3--2) and (2--1) lines. In addition, the $^{12}$CO(1--0) line is broader than that observed in the two higher transitions, and its central velocity is more blue-shifted than the other lines. RN~40 is the massive end of a finger-like extension connected to the shell. In this case one expects that the emission from the connecting filament is blue-shifted relative to the head, because the massive core has been lagging behind in the acceleration of the shell, just as for elephant trunks (Gahm et al. \cite{gahm06}). 

\subsection{Background components}
\label{sec:background}

\begin{figure}[t]
\centering
\includegraphics[angle=00, width=8cm]{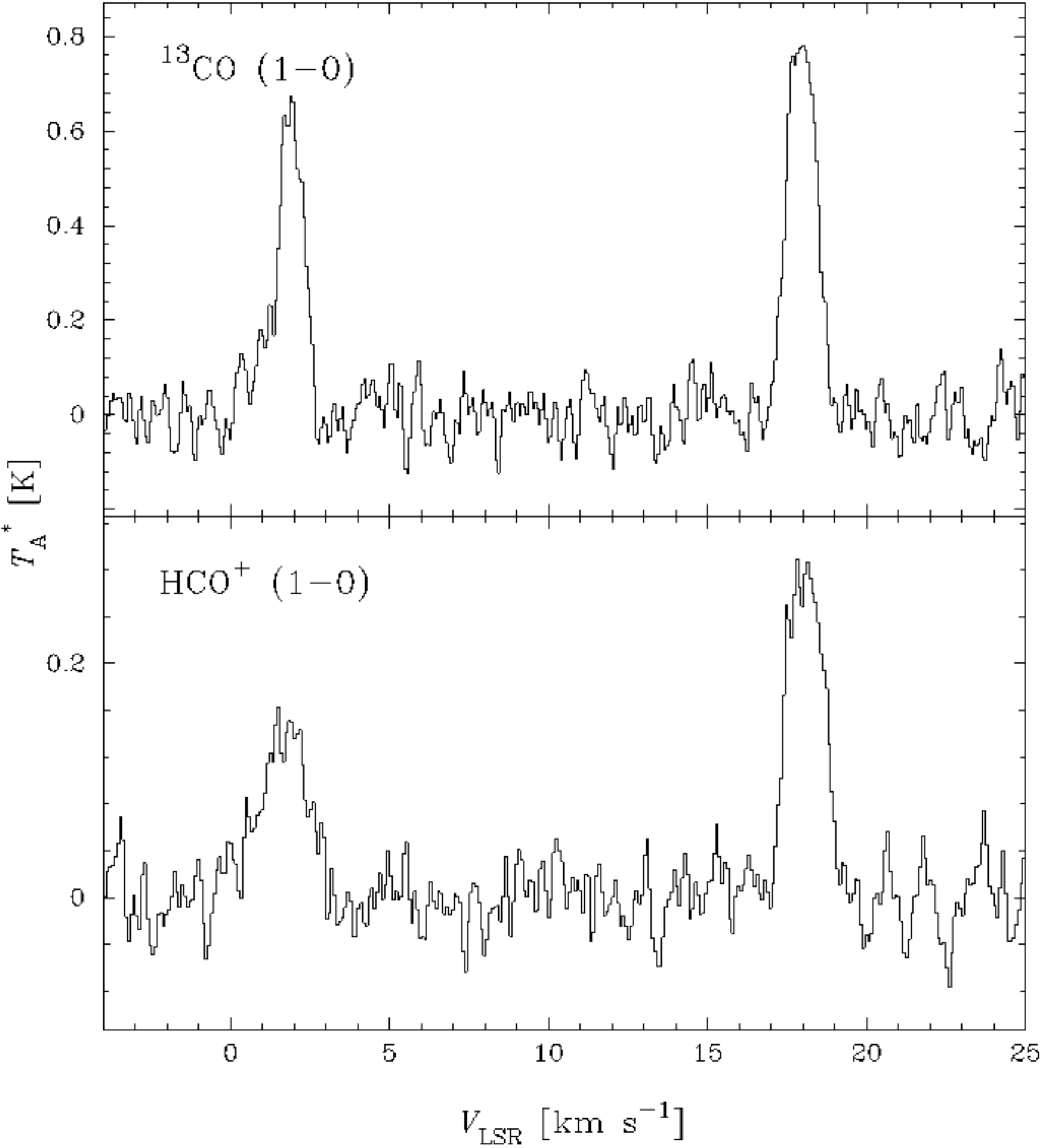}
\caption{\mbox{$^{13}$CO(1--0)} (top) and HCO$^{+}$(1--0) (bottom) spectra from the central position of RN~5. The globulette signal is at +2~km\,s$^{-1}$, and the component at +18~km\,s$^{-1}$ comes from the background shell.}
\label{fig:background}
\end{figure}

Seven globulettes show significantly red-shifted background components. An example of this is shown in Fig.~\ref{fig:background} with the $^{13}$CO and HCO$^{+}$ spectra obtained towards the central position of RN~5. The background component at $\sim$ +18~km\,s$^{-1}$ has about the same spectral appearance over our nine-grid map. Most background components have velocities of around +17~km\,s$^{-1}$, and in the maps in Dent et~al. (\cite{dent09}) this red-shifted gas extends over large areas in the upper part of the Rosette Nebula. This component is commonly identified with gas at the remote side of the nebula. 

It is then puzzling that since the central cluster has a radial velocity of +17~km\,s$^{-1}$,  a remote shell in a spherical, expanding system is expected to move at $\sim$~+34~km\,s$^{-1}$, and not at the cluster velocity. If the +17~km\,s$^{-1}$ component indeed comes from a remote shell, one would infer that the nebula is not symmetric, but shaped more like a cone, a "champagne flow" extending from the base of the nebula at the border of the giant molecular cloud extending to the southeast of the cluster. 

However, on radio continuum maps of the Rosette complex provided in Graham  (\cite{graham82}) the \ion{H}{ii} region appears as a very round object indeed. Moreover, the central velocities of the radio recombination lines of H~110$\alpha$, He~112$\alpha$, and H~112$\alpha$ all fall around +17~km\,s$^{-1}$ (Celnik\cite{celnik85}; Viner et~al. \cite{viner79}). As referred to above, the corresponding line widths range from 15 to 31~km\,s, which is more consistent with a spherically expanding system.  There are no foreground obscuring clouds in the area in question, so the implication appears to be that the molecular background emission comes from an extended shell that is located in the middle of the \ion{H}{ii} region. Alternatively, if the emission does come from a shell bordering the remote side of the nebula, then this shell has not yet reached its final expansion velocity. 

\subsection{Nature of globulettes}
\label{sec:origin}

We conclude that on the whole the masses estimated above from our molecular line observations with APEX are in the same range as those obtained from the NOT H$\alpha$ survey in Paper~I. This agreement is notable, since the former mass relates to the amount of gas in the globulettes, while the latter mass relates to the amount of dust, where a gas-to-dust ratio of 100 was assumed. The globulettes appear as well-confined objects of relatively uniform density, dropping sharply close to the surface. Our \J\ and P$\beta$ images show that some globulettes in addition contain very dense cores, as evidenced also  by the high extinction found in one direction through Rn 35 (Sect. \ref{sec:nir_mass}).  

In the model simulations by e.g. Kuutmann (\cite{kuutmann07}) and Henney et al. (\cite{henney09}) any clump exposed to radiation from central stars in a nebulae will develop a photoionized layer on the side facing the OB stars, and also a tail of molecular gas driven in the opposite direction. It is then puzzling that none of our selected objects shows any sign of bright rims in H$\alpha$, with the exception of RN~D, which is surrounded by a circular halo consistent with its location in the foreground to the cluster stars. However, the larger globulettes in the Rosette Nebula have bright rims in P$\beta$ and pronounced tails. Some smaller globulettes also show faint P$\beta$ emission extending over their surfaces, which may flag the presence of bright rims on the remote sides of the globulettes. Such rims would be entirely obscured in the H$\alpha$ images. Since the globulettes are all located on our side of the complex, the side exposed to radiation is partly hidden from us, and elephant trunks in the area show clear evidence of remote rims as well in H$\alpha$. The observed rims are much thinner than those encountered in the model simulations, which may be a consequence of the high density encountered already below the surface. Such thin rims may escape detection in the H$\alpha$ images, especially since the background emission is high. The thin layer of \Htwo\ fluorescent emission bordering some globulettes also indicates that the gas is dense and in molecular form just below the surface (Sect.~\ref{sec:sofi}). The global evolution of giant H II regions has been modelled in Arthur et al. (\cite{arthur11}) and has been followed up more recently in W. Henney and collaborators (Henney 2013, private communication), showing in more detail how small-scale cold clumps can erode from pillars and shells. It appears that these clumps have much in common with the globulettes studied here, and interestingly, the model clumps can be seen with or without a bright rim, depending on the viewing angle. 

We have concluded that the velocity pattern observed for the globulettes, which are distributed over a large area in the Rosette Nebula, is very smooth. The fact that the globulettes move with velocities similar to the inner remnant shell in the foreground of the nebula strongly suggests that the globulettes once were detached from eroding structures in the shell. Several objects are still connected by thin filaments to larger shell structures, and as the larger elephant trunks, they obtain this form since the more massive clump lags behind in the general expansion of the shell. We envision that in a next phase the thin filaments will erode and disappear, and the clumps will become isolated globulettes. The string of globulettes east of the Claw may have formed recently from a filament, that became detached from this rotating complex. Other quite isolated globulettes, far from any larger block of molecular gas and dust, must have been separated from larger blocks long ago. 

We therefore expect a range of ages among the globulettes as counted from the moment of detachment. There is no indication that some objects were accelerated even more since they detached, or that they were ejected from trunks and shells by some mechanism. If globulettes were subject to acceleration from the rocket effect described in Sect.~\ref{sec:intro}, one would expect a much larger spread in radial velocity among the objects and a systematic blue-shift relative to the shells by 5 to 10~km\,s$^{-1}$.

Possibly, many of the round isolated globulettes represent denser cores that were once encapsulated in less dense shells that since then have evaporated and lost most of the signatures of bright rims and tails. Physical parameters for the expanding plasma in the area of the northwestern ring were extracted in Gahm et~al. (\cite{gahm06}) from the studies of radio lines cited in Sect.~\ref{sec:velocities}, and it was found in Paper~I that the external pressure acting on the globulettes is comparable with the internal gas pressure. These external forces from the surrounding warm and turbulent medium help to confine the globulettes, and from simple virial arguments it was concluded that most objects could be gravitationally unstable and may form free-floating planetary-mass objects or brown dwarfs, which will shoot out into galactic space with velocities of  $\sim$~20~km\,s$^{-1}$. In the model simulations of Kuutmann (\cite{kuutmann07}), based on sizes and densities from Paper~I, this scenario is less evident, at least so for the smaller globulettes. A new set of calculations is warranted with more advanced cooling functions than used before. Our discovery of denser cores inside some of the larger globulettes will alter the conditions assumed for their evolution. Obviously, such dense cores can be in a state of collapse.

\section{Conclusions}
\label{sec:conclusions}

Globulettes in the northwestern and northern part of the Rosette Nebula were observed for emission from the lowest transitions in $^{12}$CO, $^{13}$CO and HCO$^{+}$, and a sample was observed also for higher transitions in CO. Masses and densities of the objects were estimated from a model assuming that the globulettes have cool and dense cores surrounded by warm but dense envelopes. In addition, NIR broad-band and narrow-band Paschen $\beta$ and \Htwo\ images were collected of fields containing globulettes. From this survey we conclude the following:

* The models were able to provide estimates only of the maximum and minimum mass of each globulette. The average masses range from about 50 to 500 $M_{J}$, and are similar to those derived in Paper~I based on column densities of dust. 

* The density inside globulettes is high, $n_{H}$~$\sim$~10$^{4}$~cm$^{-3}$, also close to surface, where the gas is in molecular form, and give rise to thin layers of fluorescent \Htwo\ emission. Thin bright rims were detected also both in P$\beta$, which might escape detection in the optical due to background emission or because they are partly hidden on the remote sides. Several globulettes are very opaque, also in the NIR, and contain dense cores.

* The internal motions are weak, but some objects eject gas tails directed away from the central cluster. The globulettes are distributed over an area spanning 13 pc in projected distance.  They all move with similar velocities, +0.5~$\pm$~2.1~km\,s$^{-1}$. With an inclination of 40$\degr$ the entire system of globulettes, shells, and trunks expands  at velocities of $\sim$~22~km\,s$^{-1}$ from the central cluster. There is no indication that globulettes are subject to additional acceleration from the rocket effect. Line components connected to a remote shell were observed at several positions at velocities $\sim$~+17~km\,s$^{-1}$. These velocities are inconsistent with a spherically symmetric expanding \ion{H}{ii} region, and other geometries and expansion scenarios must be considered.

* Extinction for stars in the NIR fields were measured for stars in the direction of globulettes and shell structures, and for stars in open areas. No evidence of objects embedded in globulettes was found.

The globulettes investigated are in different evolutionary phases. We found evidence that they detach as clumps from elephant trunks and shell structures and lag behind in the general expansion of the molecular shell. From the interaction with stellar light and the surrounding plasma the objects develop tails and bright rims and erode further to form isolated, roundish objects of high density. Such cores may collapse to form planetary-mass objects or brown dwarfs, which would shoot out into interstellar space at high velocity.

Molecular line observations with an antenna, such as the interferometer ALMA, would provide key information on the conditions prevailing in the interior of globulettes that would be of significant importance to our discussion of their fate. Finally, more detailed comparisons between current model simulations of small-scale molecular structures evolving in \ion{H}{ii} region would be of great interest and can now be realised.

\begin{acknowledgements}

This work was supported by the Magnus Bergvall Foundation. M.M. acknowledges the support from the Finnish Graduate School in Astronomy and Space Physics. L.H. acknowledges the support from the Finnish Ministry of Education project ÓUtilizing FinlandÕs membership in the European Southern ObservatoryÓ and the Academy of Finland under grant 132291. We are most grateful to the referee, William Henney, for some very valuable comments. 
 
\end{acknowledgements}

\Online

\begin{appendix}

\section{Spectral atlas}
\label{AppendixA}

\subsection{Spectra collected at APEX and Onsala}
\label{AppendixA1}
  
\begin{figure*}[t]
\centering

\includegraphics[angle=00, width=14cm]{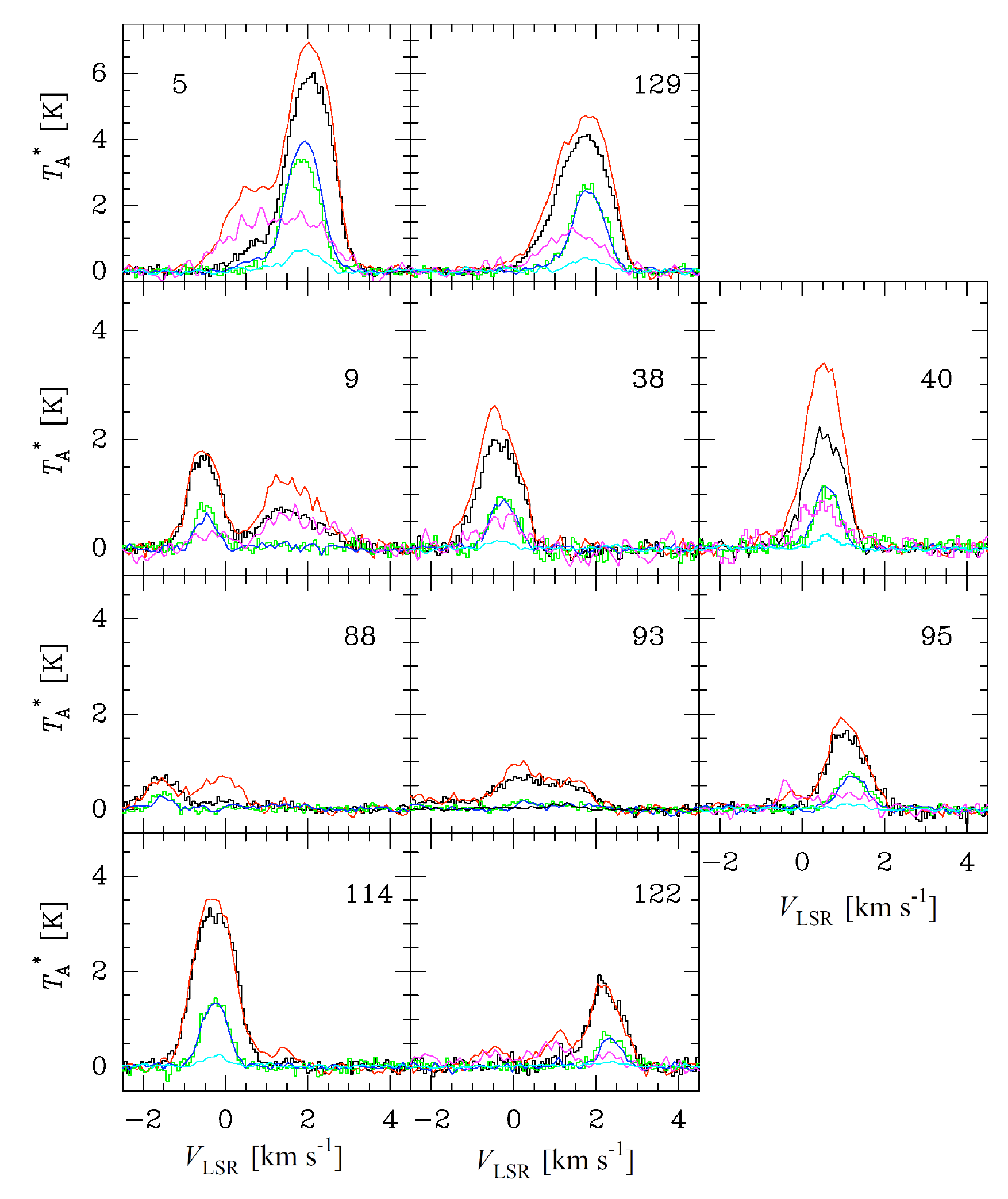}
\caption{CO spectra obtained for on-source-positions of the globulettes observed at  both at APEX and Onsala in the $T_\mathrm{A}^*$ scale. The \twco\ (3--2),(2--1) and (1--0) spectra are plotted in black, red and magenta, respectively. For the \thco spectra the colours are green, blue and cyan. The \twco and \thco (3--2) spectra are plotted as histograms. Note the different temperature scale in the uppermost row. }
\label{fig:AppendixA1}
\end{figure*}

\subsection{Spectra collected at Onsala}
\label{appendixA2}

\begin{figure*}[t]
\centering
\includegraphics[angle=00, width=8cm]{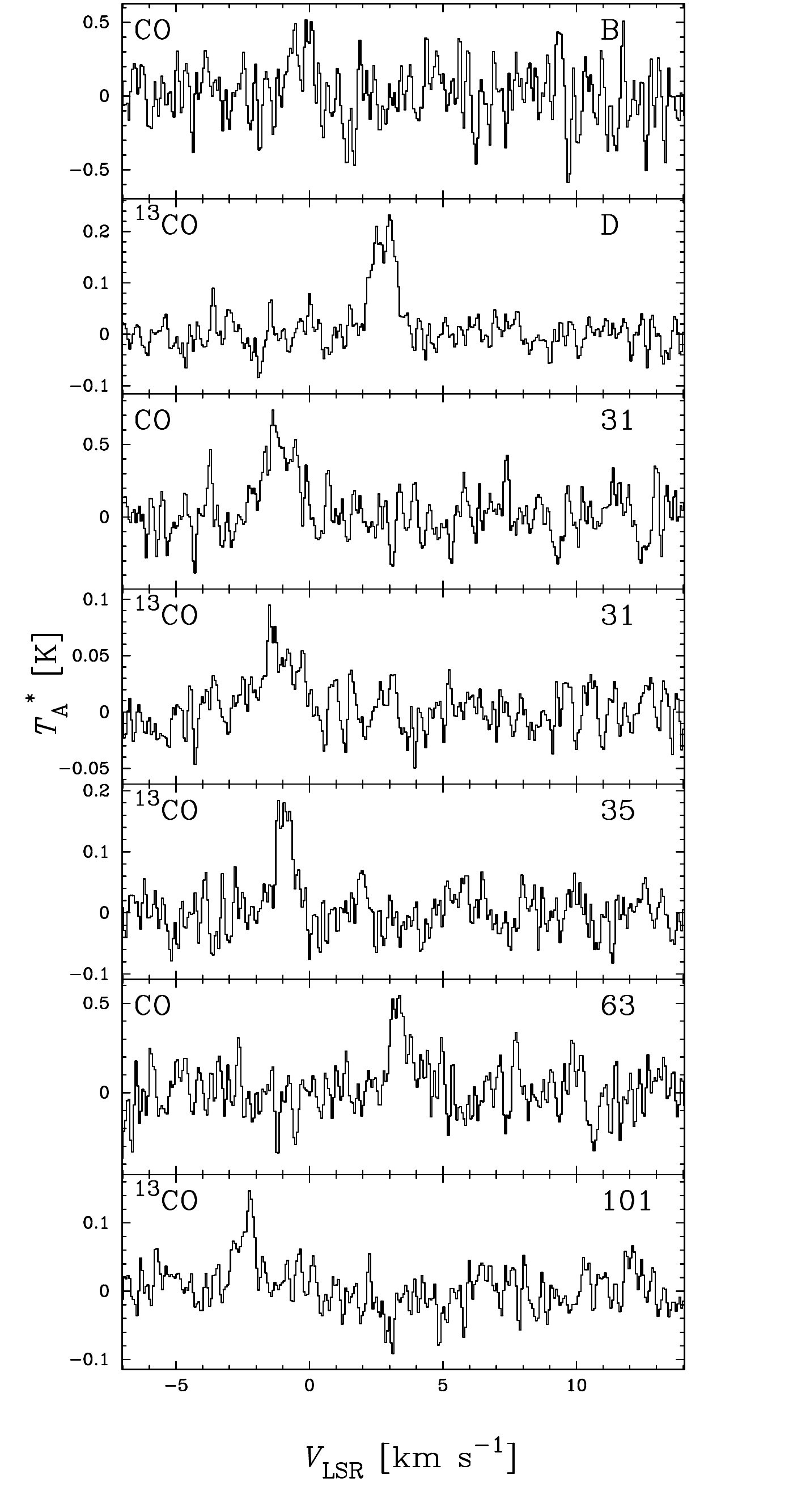}
\includegraphics[angle=00, width=7.7cm]{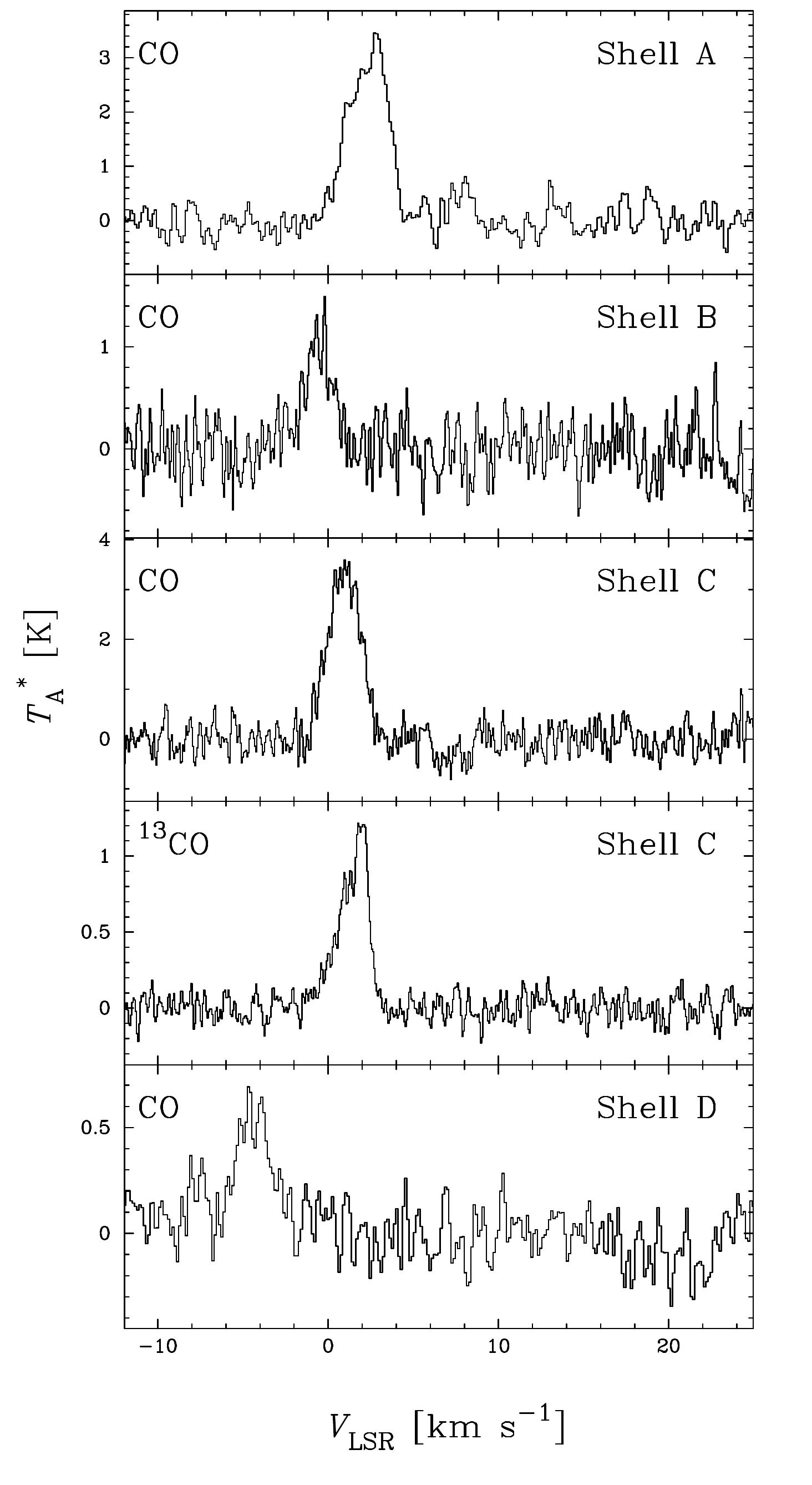}
\caption{Left: CO spectra obtained for on-source-positions of the globulettes observed only at  Onsala in the $T_\mathrm{A}^*$ scale. Right: CO spectra obtained for the selected positions in shells.} 
\label{fig:AppendixA2}
\end{figure*}

\end{appendix}

\end{document}